\documentclass[fleqn,usenatbib,useAMS]{mnras}

\usepackage{graphicx, color}
\usepackage{mathtools}

\usepackage{txfonts}

\usepackage[T1]{fontenc}
\usepackage{ae,aecompl}

\title[Supermassive Stars at the Onset of Collapse]{Maximally Rotating
  Supermassive Stars at the Onset of Collapse: Effects of
  Gas Pressure}
\author[Dennison, Baumgarte \& Shapiro]{Kenneth A.~Dennison,$^{1}$ 
Thomas W.~Baumgarte,$^{1}$ 
and Stuart L.~Shapiro$^{2,3}$ \\
 $^{1}$Department of Physics and Astronomy, Bowdoin College, Brunswick, ME 04011\\
 $^{2}$Department of Physics, University of Illinois at Urbana-Champaign, Urbana, IL 61801\\
 $^{3}$Department of Astronomy and NCSA, University of Illinois at Urbana-Champaign, Urbana, IL 61801
}

\date{Accepted XXX. Received YYY; in original form ZZZ}

\pubyear{2019}

\begin{document}
\label{firstpage}
\pagerange{\pageref{firstpage}--\pageref{lastpage}}
\maketitle

%
\begin{abstract}
The ``direct collapse" scenario has emerged as a promising
evolutionary track for the formation of supermassive black holes early
in the Universe.  In an idealized version of such a scenario, a
uniformly rotating supermassive star spinning at the mass-shedding (Keplerian) 
limit collapses gravitationally after
it reaches a critical configuration.  Under the assumption that the gas is
dominated by radiation pressure, this critical configuration is
characterized by unique values of the dimensionless parameters $J/M^2$
and $R_p/M$, where $J$ is the angular momentum, $R_p$ the polar radius,
and $M$ the mass.  Motivated by a previous perturbative treatment we
adopt a fully nonlinear approach to evaluate the effects of gas
pressure on these dimensionless parameters for a large range of
masses.  We find that gas pressure has a significant effect on the critical
configuration even for stellar masses as large as $M \simeq 10^6
M_{\odot}$.  We also calibrate two approximate treatments of the gas
pressure perturbation in a comparison with the exact treatment, and find that one
commonly used approximation in particular results in increasing
deviations from the exact treatment as the mass decreases, and the
effects of gas pressure increase.  The other approximation, however, proves to be
quite robust for all masses $M \gtrsim 10^4 M_{\odot}$.
\end{abstract}
%
%
%
\section{Introduction}
\label{sec:intro}
%

Supermassive black holes (SMBHs) reside at the centers of galaxies.
The most recent observational confirmation was provided by the
spectacular images of the Event Horizon Telescope Collaboration
\citep[see][as well as several follow-up publications]{EHT_I_19},
showing radiation emitted by material accreting onto the SMBH at the
center of the galaxy M87 and shadowing by the black hole's event horizon.  
Accreting SMBHs are also believed to power
quasars and active galactic nuclei, which have been observed out to
large cosmological distances \citep[see, e.g.,][]{Fan06,Fanetal06}.
Examples of quasars at large distances include J1342+0928, at a
redshift of $z \simeq 7.5$, and powered by a SMBH with mass of
approximately $7.8 \times 10^8 M_\odot$ \citep{Banetal17}, J1120-0641,
at a redshift of $z \simeq 7.1$ and with a black-hole mass of
approximately $2.0 \times 10^9 M_{\odot}$ \citep{Moretal11}, as well
as the ultra-luminous quasar J0100+2802 at a redshift of $z = 6.3$ and
with a mass of about $1.2 \times 10^{10} M_\odot$ \citep{Wuetal15}.
The existence of such massive black holes at so early an age in the Universe
poses an important question \citep[see, e.g.,][for
  reviews]{Sha04,Hai13,LatF16,SmiBL17} -- namely, how could they have
formed in such a short time?


One possible evolutionary scenario involves the collapse of
first-generation -- i.e.~Population III (Pop III) -- stars to form
seed black holes, which then grow through accretion and/or mergers.
Growth by merger may be limited by recoil speeds \citep{Hai04}.  Growth by
accretion depends in part on the efficiency of the conversion
of matter to radiation, and is usually limited by the Eddington luminosity \citep{Sha05,PacVF15}.
While this already constrains the formation of SMBHs from stellar-mass black holes \citep{SmiBL17}, 
including the effects of photoionization and heating appears to reduce the accretion 
rate to just a fraction of the Eddington limit (see \citet{AlvWA09,MilBCO09}; see 
also \citet{WhaF12} for how natal kicks affect the accretion rate,  
as well as \citet{SmiRDNOW18} for recent simulations 
in a cosmological context).  It is difficult to see, 
therefore, how seed black holes with masses of
Pop III stars, about 100 $M_{\odot}$, could grow to the masses of
SMBHs by $z \simeq 7$.  In fact, \citet{Banetal17} argue that the
existence of the objects J1342+0928, J1120-0641, and J0100+2802 ``is
at odds with early black hole formation models that do not involve
either massive ($\gtrsim 10^4 M_{\odot}$) seeds or episodes of
hyper-Eddington accretion" (see also their Fig.~2).  The observation
of these distant quasars therefore suggests the direct collapse of
objects with masses of $M \gtrsim 10^{4-5} M_{\odot}$ as a plausible
alternative scenario for the formation of SMBHs
\citep[e.g.][]{Ree84,LoeR94,OhH02,BroL03,KouBD04,Sha04,LodN06,BegVR06,RegH09b,Beg10,AgaKJNDVL12,JohWLH13}.

The progenitor object in such a ``direct collapse" scenario is often
referred to as a supermassive star (SMS).  The properties of SMSs have
been the subject of an extensive body of literature (see, e.g.,
\citet{Ibe63,HoyF63,Cha64,BisZN67,Wag69,AppF72,BegR78,FulWW86} for
some early references, as well as \citet[][hereafter ST]{ShaT83}, 
\citet{ZelN71}, and \citet{KipWW12} for textbook treatments).  
Numerous authors and groups have studied possible avenues for their
formation (see, e.g.,
\citet{SchPFGL13,HosYIOY13,SakHYY15,UmeHOY16,WooHWHK17,HaeWKHW18a,HaeWKHW18b};
see also \citet{WisRONDX19} for recent simulations in the context of
cosmological evolutions) as well as their ability to avoid
fragmentation \citep[e.g.][and references
  therein]{BroL03,WisTA08,RegH09a,LatSSN13,VisHB14,MayFBQRSW15,SunRS19}.

In \citet[][hereafter Paper I]{BauS99b}, we considered an
idealized evolutionary scenario for rotating SMSs.  We assumed that
SMSs are dominated by radiation pressure, and that they cool and
contract while maintaining uniform rotation.  Since the star will spin
up during the contraction, it will ultimately reach mass-shedding,
i.e.~the Kepler limit, and will subsequently evolve along the
mass-shedding limit \citep[see also][]{BauS99a}.  Ultimately, the SMS
reaches a critical configuration at which it becomes radially unstable to
collapse to a black hole.  The critical configuration is characterized
by unique values of the dimensionless parameters $J/M^2$ and $R_p/M$,
where $J$ is the angular momentum, $R_p$ the polar radius, $M$ the
mass, and where we have adopted geometrized units with $G = c = 1$.
We computed the values of these parameters both from numerical models
of fully relativistic, rotating $n = 3$ polytropes, and from an
approximate but analytical energy functional approach that accounts
for the stabilizing effects of rotation and the destabilizing effects
of relativistic gravity with leading-order terms only.  Both
approaches result in similar values for the critical parameters (see
Table 2 in Paper I).

The uniqueness of the parameters characterizing the critical
configuration implies that the subsequent evolution, namely the
collapse to a black hole, as well as the gravitational wave signal
emitted in the collapse, is unique as well.  Numerical simulations
have shown that this collapse will lead to a spinning black hole with
mass $M_{\rm BH} / M \simeq 0.9$ and angular momentum $J_{\rm
  BH}/M_{\rm BH}^2 \simeq 0.7$, surrounded by a disk with mass $M_{\rm
  disk}/M \simeq 0.1$ \citep[see,
  e.g.,][]{ShaS02,ShiS02,LiuSS07,MonJM12,ShiSUU16,UchSYSU17,SunPRS17,SunRS18}.

Given the importance of the critical parameters, we examined in
\citet[][hereafter Paper II]{ButLBS18} to what degree they depend on
some of the assumptions made, and computed leading-order corrections
due to gas pressure, magnetic fields, dark matter and dark energy.  We
determined these corrections using a perturbative treatment based on the
energy functional approach mentioned above.  As one might expect, the
largest corrections by far are those caused by gas pressure.  We
treated the effects of gas pressure using two different
approximations: one based on a formal expansion (``Approximation I",
see Section 17.3 in ST, as well as Section \ref{sec:eos:approachI}
below), and the other by adjusting the polytropic index $n$
(``Approximation II", see, e.g., Exercise 17.3 in ST, Problem 2.26 in
\citet{Cla83}, as well as Section \ref{sec:eos:approachII} below).
The latter approach, Approximation II, is very simple to implement,
and is therefore quite commonly used in numerical simulations
\citep[see, e.g.,][for recent examples]{ShiUS16,SunRS18}.  While it
results in expressions for the non-dimensional parameters discussed
above that are identical to those from Approximation I, at least to
leading order, expressions for some dimensional quantities differ even
at leading order.

Motivated by this observation, we revisit in this paper the effects of
gas pressure on maximally rotating SMSs at the onset of collapse.  We
improve on our treatment in Paper II in two ways.  First, we use the
``Rotating neutron star" (RNS) code of \cite{SteF95} to construct
fully relativistic models of rotating SMSs, rather than relying on a
perturbative treatment within the energy functional approach.  Second,
we employ exact expressions for a mixture of radiation and gas
pressure in addition to the two approximate treatments of gas pressure
described above.  As a result, we can treat these stars accurately
even for less massive models, for which the gas pressure becomes
increasingly important, and can calibrate the accuracy of the two
approximate treatments and their impact on this idealized
direct-collapse scenario.   We note, though, that we ignore other
effects that may become important for smaller masses, including 
electron-positron pair production or nuclear reactions.  Our findings are 
summarized in Fig.~\ref{polarradandangmomfig} 
below, which shows the dimensionless
parameters $R_p/M$ and $J/M^2$ for the critical configuration as a
function of stellar mass for a large range of stellar masses.  We find
good agreement between the exact and approximate treatments of the gas
pressure, as well as with the perturbative results of Paper II, for
large masses with $M \gtrsim 10^6 M_{\odot}$.  This confirms our
finding of Paper II that, even for these large masses, gas pressure
has an important effect on the above parameters.  For smaller masses
both approximations lead to deviations from the exact treatment of 
gas pressure, but those stemming from Approximation II are significantly larger
than those from Approximation I.

This paper is organized as follows.  In Section \ref{sec:eos} we
derive the equation of state (EOS) for a SMS supported by a
combination of radiation and gas pressure.  We model this EOS in three
different ways: exactly, assuming that the star is isentropic (Section
\ref{sec:eos:exact}), as well as using the two Approximations I and II
described above (Sections \ref{sec:eos:approachI} and
\ref{sec:eos:approachII}).  In Section \ref{sec:nr} we use these three
treatments of the EOS to explore their effects on equilibrium models
of nonrotating, spherically-symmetric SMSs.  In Section \ref{sec:rot}
we consider rotating SMSs and determine the parameters characterizing
their critical configurations at the onset of collapse to a black hole.
We conclude in Section \ref{sec:sum} with a brief summary.

%
\section{Equation of State}
\label{sec:eos}
%

In this Section we use thermodynamic relationships to derive the EOS
for a SMS supported by both radiation and gas pressure.
We first treat the gas pressure terms exactly, assuming that the star
is isentropic, and then introduce two different
approximations.\footnote{We closely follow the treatment of Paper II
  in this discussion.} We close this Section with a description of our
numerical implementation of the different EOSs.

\subsection{Exact Approach to Handling Gas Pressure}
\label{sec:eos:exact}

We begin by finding expressions for the total pressure, total internal
energy density, and total entropy per baryon. We then introduce
dimensionless variables, collect the key expressions, and discuss our
approach to generating a tabulated EOS, leaving the numerical details
to Section \ref{sec:eos:numerical}.

\subsubsection{Total Pressure}
\label{sec:eos:exact:tp}

The total pressure, $P$, is the sum of the radiation and gas
pressures,
\begin{equation}
\label{Ptot1}
P = P_{r} + P_{g}.
\end{equation}
The radiation pressure $P_r$ is given by
\begin{equation}
\label{Prad}
P_{r} = \frac{1}{3}aT^{4},
\end{equation}
where $T$ is the temperature and
\begin{equation}
\label{adef}
a \equiv \frac{8\pi^{5}k_{B}^{4}}{15h^{3}}
\end{equation}
the radiation constant in geometrized units.  In (\ref{adef}) we have
also introduced the Boltzmann constant $k_{B}$ and Planck's constant
$h$.

Assuming a fully ionized hydrogen gas for simplicity, the gas pressure
$P_{g}$ is
\begin{equation}
\label{Pgas}
P_{g} = 2n_{B}k_{B}T,
\end{equation}
where
\begin{equation}
\label{numdens}
n_{B} = \frac{\rho_{0}}{m_{B}}
\end{equation}
is the baryon number density, $\rho_{0}$ is the rest-mass density, and
$m_{B}$ is the baryon rest mass.  The total pressure is then given by
\begin{equation}
\label{Ptot2}
P = P_{r} + P_{g} = \frac{1}{3}aT^{4} + 2n_{B}k_{B}T.
\end{equation}

\subsubsection{Total Internal Energy Density}
\label{sec:eos:exact:tied}

Similarly, the total internal energy density $\epsilon$ is the sum of
contributions from the radiation,
\begin{equation}
\label{Erad}
\epsilon_{r} = aT^{4},
\end{equation}
and the (nonrelativistic) plasma,
\begin{equation}
\label{Egas}
\epsilon_{g} = 3n_{B}k_{B}T,
\end{equation}
where we have again assumed a fully ionized hydrogen gas.  We then
have
\begin{equation}
\label{Etot2}
\epsilon = \epsilon_{r} + \epsilon_{g} = aT^{4} + 3n_{B}k_{B}T.
\end{equation}
The total (energy) density $\rho$ is the sum of the rest-mass density
and the total internal energy density, i.e.
\begin{equation}
\label{densities}
\rho = \rho_{0} + \epsilon.
\end{equation}

\subsubsection{Total Entropy per Baryon}
\label{sec:eos:exact:tepb}

The total entropy per baryon, $s$, is again the sum of contributions
from the radiation and the gas,
\begin{equation}
\label{Stot1}
s = s_{r} + s_{g},
\end{equation}
and is related to the internal energy density and pressure through the
first law of thermodynamics,
\begin{equation}
\label{firstlaw}
Tds = d\left(\frac{\epsilon}{n_{B}}\right)+Pd\left(\frac{1}{n_{B}}\right).
\end{equation}
The photon entropy per baryon, $s_{r}$, is
\begin{equation}
\label{Srad}
s_{r} = \frac{4am_{B}T^{3}}{3\rho_{0}},
\end{equation}
and the gas entropy per baryon, $s_{g}$, is
\begin{equation}
\label{Sgas}
s_{g} = k_{B}\ln\left(\frac{4m_{e}^{3/2}m_{B}^{7/2}}{\rho_{0}^{2}}\left(\frac{k_{B}T}{2\pi\hbar^{2}}\right)^{3}\right) + 5k_{B},
\end{equation}
where $m_{e}$ is the electron mass and $\hbar\equiv h/2\pi$.
Substituting eqs. (\ref{Srad}) and (\ref{Sgas}) into
eq. (\ref{Stot1}), we find that the total entropy per baryon is
\begin{equation}
\label{Stot2}
s = \frac{4am_{B}T^{3}}{3\rho_{0}} + k_{B}\ln\left(\frac{4m_{e}^{3/2}m_{B}^{7/2}}{\rho_{0}^{2}}\left(\frac{k_{B}T}{2\pi\hbar^{2}}\right)^{3}\right) + 5k_{B}.
\end{equation}

\subsubsection{Collecting Equations}
\label{sec:eos:exact:collect}

In geometrized units the pressure and the various energy densities all
have the same units of $length^{-2}$.  Therefore, we can
nondimensionalize them using the same constant, which proves to be
convenient for later numerical work.  Defining a constant $K$ with
units of $length^{2/3}$,
\begin{equation}
\label{Kdef}
K \equiv \frac{a}{3}\left(\frac{3s}{4m_{B}a}\right)^{4/3},
\end{equation}
we define dimensionless pressure, rest-mass density, internal energy
density, and total density as
\begin{align}
\bar{P} & \equiv K^{3} P,  \label{dimPdef} \\
\bar{\rho_0} & \equiv K^{3} \rho_0, \label{dimrho0def} \\
\bar{\epsilon} & \equiv K^{3}\epsilon \label{dimepsdef} ,
\end{align}
and
\begin{equation}
\label{dimrhodef}
\bar{\rho} \equiv K^{3}\rho,
\end{equation}
respectively.  In terms of these dimensionless variables,
eqs.~(\ref{Ptot2}), (\ref{Etot2}), (\ref{densities}), and
(\ref{Stot2}) now take the form
\begin{align}
\label{dimPequ}
\bar{P} & = \frac{1}{3}aK^{3}T^{4} + 2\frac{\bar{\rho_{0}}}{m_{B}}k_{B}T, \\
\label{dimepsilonequ}
\bar{\epsilon} & = aK^{3}T^{4} + 3\frac{\bar{\rho_{0}}}{m_{B}}k_{B}T, \\
\label{dimrhoequ}
\bar{\rho} & = \bar{\rho_{0}} + aK^{3}T^{4} + 3\frac{\bar{\rho_{0}}}{m_{B}}k_{B}T,
\end{align}
and
\begin{equation}
\label{sequ}
s = \frac{4a}{3}\frac{T^{3}}{\bar{\rho_{0}}}K^{3}m_{B} + k_{B}\ln\left(\frac{4m_{e}^{3/2}m_{B}^{7/2}}{\bar{\rho_{0}}^{2}}K^{6}\left(\frac{k_{B}T}{2\pi\hbar^{2}}\right)^{3}\right) + 5k_{B}.
\end{equation}

Given a pressure $\bar{P}$ and an entropy per baryon $s$, we solve
eqs.~(\ref{dimPequ}) and (\ref{sequ}) simultaneously for $T$ and
$\bar{\rho_{0}}$, which we then substitute into
eqs.~(\ref{dimepsilonequ}) and (\ref{dimrhoequ}) to calculate
$\bar{\epsilon}$ and $\bar{\rho}$, respectively.  The result is a
tabulated EOS that we use in numerical calculations in Sections
\ref{sec:nr} and \ref{sec:rot}.  We discuss the construction of these
tabulated EOSs in more detail in Section \ref{sec:eos:numerical}
below.

Instead of adopting an exact description of radiation and gas
pressure, it is also common to use approximate treatments.  We
introduce two different approximations, ``Approximation I" and
``Approximation II", in Sections \ref{sec:eos:approachI} and
\ref{sec:eos:approachII} below.  For the purpose of comparing these
approximate treatments with the exact solution it is convenient to
define a small dimensionless parameter $\beta$,
\begin{equation}
\label{defbeta}
\beta\equiv 8k_{B}/s \approx P_{g}/P_{r}.
\end{equation}
We note that slightly different definitions of $\beta$ are used in the
literature.  In Paper II, in particular, we defined $\beta$ in terms
of the radiation entropy $s_r$ rather than the total entropy $s$.  To
linear order, however, the two definitions are equivalent, so that all
linear-order expressions in Paper II can be used without modification.
With the definition (\ref{defbeta}) a constant $\beta$ now means
constant \emph{total} entropy per baryon throughout a star, instead of
constant \emph{radiative} entropy per baryon.  Constant total entropy
per baryon is the more realistic assumption, and is made plausible for
SMSs because they are expected to be convective (see, e.g. the
Appendix of \cite{LoeR94}).

\subsection{Approximation I}
\label{sec:eos:approachI}

Approximation I is based on a formal expansion, and takes into account
the effects of the gas to leading order only.  We refer the reader to
Section 17.3 in ST for a detailed treatment, but review the main
results here.  If $s_{g} \ll s_{r}$, we can approximate $s_{r}$ with
$s$ and write the temperature as
\begin{equation}
\label{TApp1}
T \approx \left(\frac{3s\rho_{0}}{4m_{B}a}\right)^{1/3}\left(1-\frac{s_{0}}{3s}-\frac{k_{B}}{3s}\ln\left(\frac{3s}{4m_{b}a\rho_{0}}\right)\right),
\end{equation}
where $s_{0}$ is defined as 
\begin{equation}
\label{defs0}
s_{0} \equiv \left(3\ln\left(\frac{k_{B}}{2\pi\hbar^{2}}\right)+\frac{3}{2}\ln m_{e}+\frac{7}{2}\ln m_{B}+2\ln 2 + 5\right)k_{B}.
\end{equation}
(see eq.~(17.3.4) in ST).  The natural scale factor, which we called
$K_{I}$ in paper II, is the same as $K$ defined in (\ref{Kdef}), $K_I
= K$.
Defining the auxiliary functions
\begin{equation}
\label{lambdaApp1def}
\bar{\lambda} \equiv -\frac{4s_{0}}{s} + \frac{12k_{B}}{s} - \frac{4k_{B}}{s}\ln\frac{3s}{4m_{B}a}
\end{equation}
and
\begin{equation}
\label{muApp1def}
\bar{\mu} \equiv \frac{4k_{B}}{s},
\end{equation}
we write the internal energy density as
\begin{equation}
\label{epsilon1App1}
\epsilon \approx K\rho_{0}^{4/3}\left(3 + \bar{\lambda} + \bar{\mu}\ln\rho_{0}\right).
\end{equation}
The functions $\bar{\lambda}$ and $\bar{\mu}$ are decorated with bars
because they are dimensionless versions of the corresponding functions
$\lambda$ and $\mu$ defined in eqs.~(17.3.11) and (17.3.12) of ST.

In terms of $\beta$, eq. (\ref{epsilon1App1}) becomes
\begin{equation}
\label{epsilon2App1}
\epsilon \approx K\rho_{0}^{4/3}\left(3 - \beta\left(1 - \frac{5}{2}\ln\beta - \frac{1}{2}\ln\left(K^{3}\rho_{0}\right)+\frac{1}{2}\ln\eta\right)\right),
\end{equation}
where
\begin{equation}
\label{etaApp1def}
\eta \equiv \frac{2^{4}3^{4}5^{2}}{\pi^{7}}\left(\frac{m_{e}}{m_{B}}\right)^{3/2} \approx 1.367\times 10^{-4}.
\end{equation}
The pressure can be found in terms of $\epsilon$ as
\begin{equation}
\label{PApp1}
P \approx \frac{1}{3}\frac{1+\beta}{1+\beta/2}\epsilon.
\end{equation}
As in Section \ref{sec:eos:exact:collect}, we define dimensionless
fluid variables using the scaling relations (\ref{dimPdef}) through
(\ref{dimrhodef}).  Given the pressure and the entropy per baryon, we
can solve eq.~(\ref{PApp1}) for the internal energy density
$\epsilon$.  Substitution into eq.~(\ref{epsilon2App1}) then allows us
to find a numerical solution for the rest-mass density $\rho_{0}$,
which can simply be added to the internal energy density to find the
total density $\rho$.  From these, we construct another tabulated EOS
(see also Section \ref{sec:eos:numerical} below).

\subsection{Approximation II}
\label{sec:eos:approachII}

A pure radiation gas is an $n = 3$, or $\Gamma = 1 + 1/n = 4/3$
polytrope.  In Approximation II, the EOS is still taken to be of
polytropic form, with the effects of the gas pressure approximated by
a small change in the polytropic index (see, e.g., Exercise 17.3 in ST,
and Problem 2.26 in \cite{Cla83}).  We compute the adiabatic exponent
from
\begin{equation}
\label{Gamma1App2}
\Gamma_{1} \equiv \left(\frac{d\ln P}{d\ln\rho_{0}}\right)_{s} = \frac{4}{3} + \frac{\beta\left(4+\beta\right)}{3\left(1+\beta\right)\left(8+\beta\right)} \approx \frac{4}{3}+\frac{\beta}{6},
\end{equation}
and require the pressure $P$ to obey
\begin{equation}
\label{PApp2}
P = K_{II}\rho_{0}^{\Gamma_{1}}.
\end{equation}
We find that $K_{II}$ is
\begin{equation}
\label{KIIApp2}
K_{II} = (1+\beta)K\rho_{0}^{-\beta/6},
\end{equation}
which is not truly constant.  Approximating $K_{II}$ as independent of
$\rho_{0}$ for small $\beta$, we can find the internal energy density
to be
\begin{equation}
\label{epsilonApp2}
\epsilon = n_{1}P,
\end{equation}
where the approximate polytropic index is
\begin{equation}
\label{n1App2}
n_{1} = \frac{1}{\Gamma_{1} -1} = \frac{3}{1+\beta/2}.
\end{equation}
The scale factor used to define dimensionless quantities is now
$K_{II}\approx (1+\beta)K$.  Given a pressure and an entropy per
baryon, we can use eq. (\ref{epsilonApp2}) to calculate the internal
energy density $\epsilon$ and eq. (\ref{PApp2}) to calculate the
rest-mass density $\rho_{0}$.  The total density $\rho$ is again the
sum of $\rho_{0}$ and $\epsilon$.

\subsection{Numerical Implementation}
\label{sec:eos:numerical}

Given an EOS, a pressure $\bar{P}$, and a total entropy per baryon
$s$, we would like to calculate the remaining thermodynamic variables.
For all three approaches, we first
compute $\beta$ from $s$ using (\ref{defbeta}).  For Approximation II
(Section \ref{sec:eos:approachII}), we can then compute all quantities
analytically.  For the exact approach (Section
\ref{sec:eos:exact:collect}) and Approximation I (Section
\ref{sec:eos:approachI}), however, we need to find roots of equations
numerically.  In practice, we use the Numerical Recipes
(\cite{PreTVF02}) routines {\tt rtsafe} and {\tt mnewt} for
one-dimensional and two-dimensional iterative root-finding,
respectively.  These routines require analytical derivatives.  For the
exact EOS, for example, we use eqs.~(\ref{dimPequ}) and (\ref{sequ})
to define
\begin{align}
F_{1}\left(T,\bar{\rho}_{0}\right)  \equiv & \bar{P} - \frac{1}{3} a K^{3} T^{4} - \frac{2\bar{\rho}_{0}k_{B}T}{m_{B}}, \\
F_{2}\left(T,\bar{\rho}_{0}\right) \equiv & s - \frac{4 a m_{B} K^{3} T^{3}}{3 \bar{\rho}_{0}} - \\
&  k_{B}\ln\left(\frac{4 m_{e}^{3/2}m_{B}^{7/2}K^{6}}{\bar{\rho}_{0}^{2}}\left(\frac{k_{B}T}{2\pi\hbar^{2}}\right)^{3}\right) - 5k_{B}.
\end{align}
To solve eqs.~(\ref{dimPequ}) and (\ref{sequ}) simultaneously for $T$ and $\bar{\rho}_{0}$, the required analytical derivatives are the Jacobian matrix elements
\begin{align}
J_{11} \equiv &  \partial_{T}F_{1} = -\frac{4}{3} a K^{3} T^{3} - \frac{2\bar{\rho}_{0}k_{B}}{m_{B}},\\
J_{12} \equiv & \partial_{\bar{\rho}_{0}}F_{1} = -\frac{2k_{B}T}{m_{B}},\\
J_{21} \equiv & \partial_{T}F_{2} = -\frac{4am_{B}K^{3}T^{2}}{\bar{\rho}_{0}}-\frac{3k_{B}}{T},\\
J_{22} \equiv & \partial_{\bar{\rho}_{0}}F_{2} = \frac{4am_{B}K^{3}T^{3}}{3\bar{\rho}_{0}^{2}}+\frac{2k_B}{\bar{\rho}_{0}}.
\end{align}
Note that $J_{21}$ is also the derivative needed for the numerical
solution of eq.~(\ref{sequ}) for $T$ when given $s$ and
$\bar{\rho}_{0}$.  In addition to derivative information, {\tt mnewt}
needs a good initial guess for the solutions $T$ and $\bar{\rho}_{0}$.
Because we expect the addition of gas terms to make only a small
difference, we can assume a polytropic solution with $n=3$ and use
\begin{align}
T_{\rm guess} & = \left(\frac{3\bar{P}}{a K^{3}}\right)^{1/4}, \\
\bar{\rho}_{0,{\rm guess}} & = \frac{4 a m_{B} K^{3} T_{{\rm guess}}^{3}}{3s}
\end{align}
as initial guesses.  Once the solutions for $T$ and $\bar{\rho}_{0}$
have been found, $\bar{\epsilon}$ and $\bar{\rho}$ can be found from
eqs.~(\ref{dimepsilonequ}) and (\ref{dimrhoequ}).  Our EOS class is
called directly from our TOV-solver for the calculations in Section
\ref{sec:nr}.  A separate driver routine calls our EOS class to
generate tabulated EOSs suitable for input to the RNS code
(\cite{SteF95}) used for calculations in Section \ref{sec:rot}.

%
\section{Nonrotating Supermassive Stars}
\label{sec:nr}
%

As a first experiment we explore the effects of the different
treatments of the EOS on the structure of nonrotating SMS.  To do so,
we solve the Tolman-Oppenheimer-Volkoff equations \citep{OppV39,Tol39}
\begin{equation}
\label{dmdrTOV}
\frac{dm}{dr} = 4\pi\left(\rho_{0}+\epsilon\right)r^{2},
\end{equation}
and
\begin{equation}
\label{dPdrTOV}
\frac{dP}{dr} = -\left(\rho_{0}+\epsilon+P\right)\frac{m + 4\pi Pr^{3}}{r^2\left(1-2m/r\right)},
\end{equation}
where $m(r)$ is the mass inside areal radius $r$.  The stellar radius $R$ is
defined as the value of $r$ at which the pressure $P$ first vanishes.
The stellar mass is then given by $M = m(R)$.  Eqs. (\ref{dmdrTOV})
and (\ref{dPdrTOV}) can be non-dimensionalized as previously discussed
for the EOSs.  For each of our three approaches to handling gas
pressure we pick a value for the total entropy per baryon and
numerically integrate the TOV equations at fixed entropy for a variety of central
rest-mass densities.  At a critical central rest-mass density
$\rho_{0,c}$ the mass $M$ of the star along a sequence of constant entropy
is maximized, marking the onset
of radial instability (``turning-point" criterion).  We call this mass $M_{\rm crit}$.  As
motivated in Paper II, we combine these critical parameters into a
single dimensionless parameter $x_{\rm crit}$
\begin{equation}
\label{defxcrit} 
x_{\rm crit} \equiv \bar{M}_{\rm crit}^{2/3}\bar{\rho}_{0,c}^{1/3} = M_{\rm crit}^{2/3}\rho_{0,c}^{1/3}.
\end{equation}

\begin{figure}
\begin{center}
\includegraphics[width= 0.47 \textwidth]{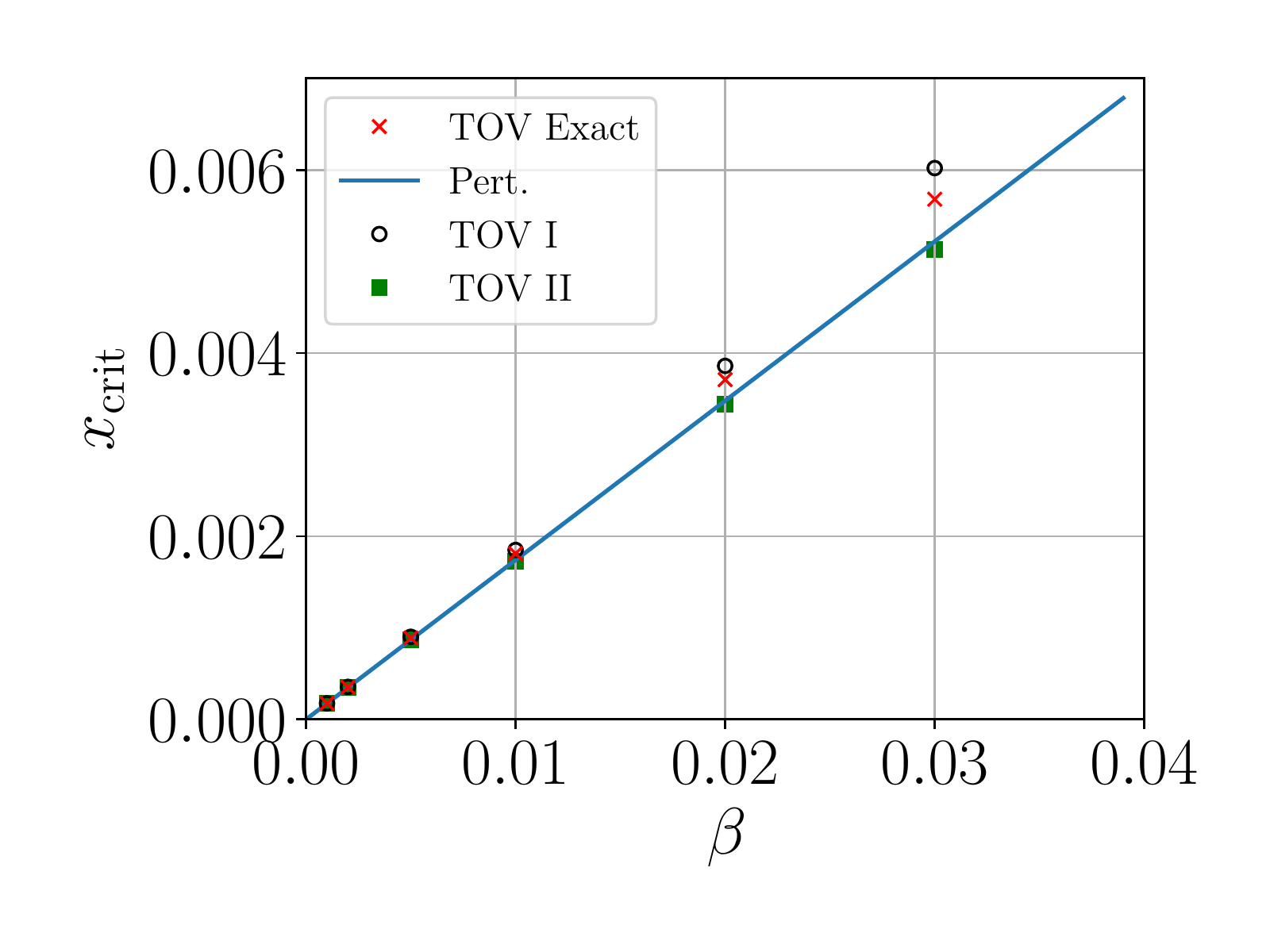}
\end{center}
\caption{The dimensionless variable $x_{\rm crit} = M^{2/3}_{\rm
    crit}\rho^{1/3}_{0, c}$ as a function of $\beta=8k_{B}/s$ for
  nonrotating SMS solutions to the TOV equations.  Crosses (red
  online) denote the numerical results for the exact treatment of the
  EOS from Section \ref{sec:nr}.  The solid line (blue online)
  represents the analytical, leading-order perturbative prediction
  (\ref{xcritmodel1}) from Section \ref{sec:nr} (which is identical
  for Approximations I and II to the EOS).  The open circles (outlined
  in black online) and filled squares (green online) denote the
  numerical results from Section \ref{sec:nr}, using Approximations I
  and II to the EOS, respectively.  For finite entropy (nonzero
  $\beta$) a SMS is partially supported by gas pressure, and nonzero
  $x_{\rm crit}$ indicates that this stabilizes it against collapse
  for central densities below $\rho_{0,c}$.  Compare with Fig.~1 of
  Paper II.  As suggested in Paper II, Approximation I is indeed
  closer to the exact solution, despite Approximation II agreeing
  better with the leading-order perturbative prediction.}
\label{xcritfig}
\end{figure}

\begin{figure}
\begin{center}
\includegraphics[width= 0.47 \textwidth]{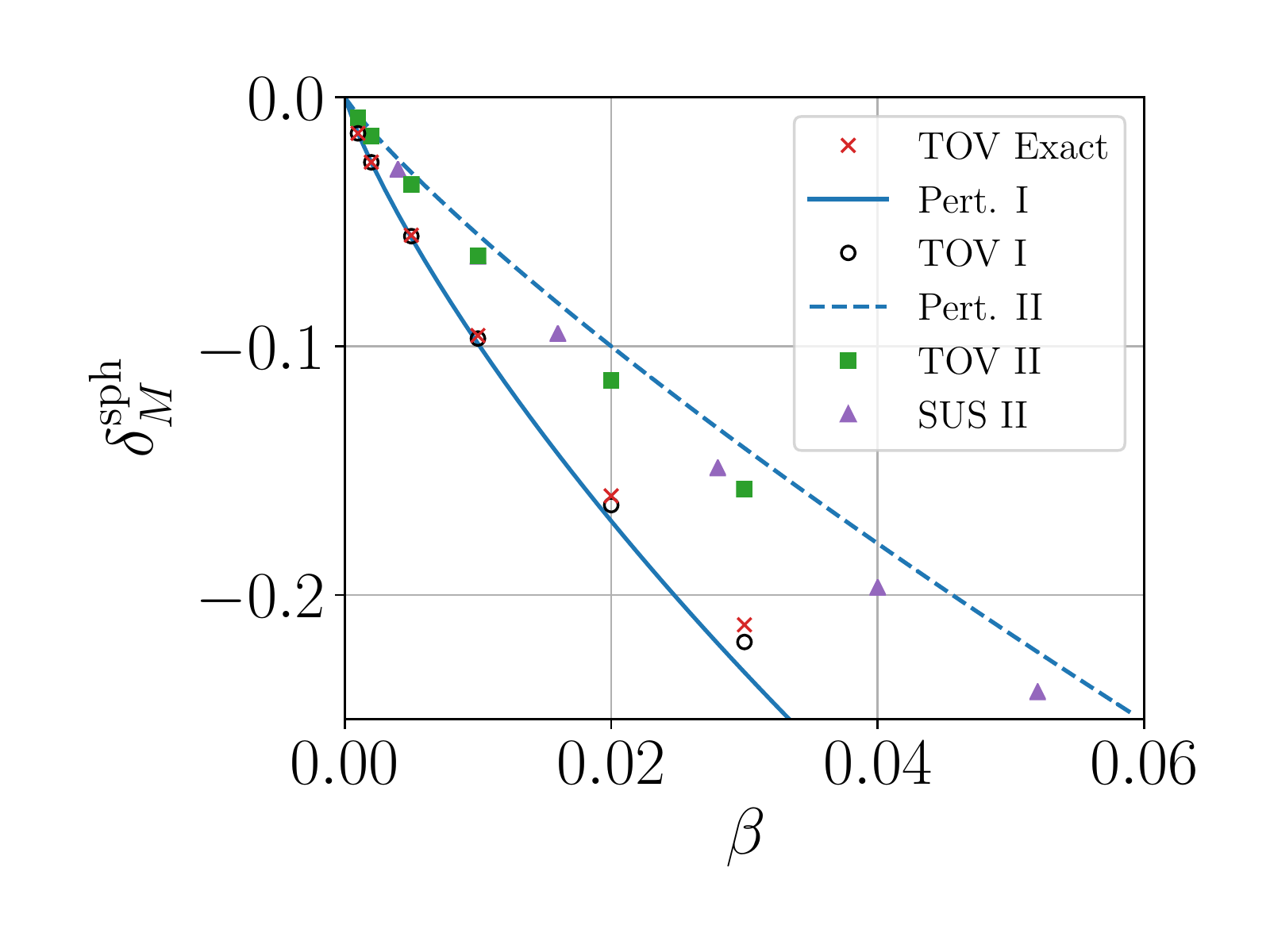}
\end{center}
\caption{The relative change in the mass $\delta^{\rm sph}_{M}$ (see
  (\ref{defdeltaM})) as a function of $\beta=8k_{B}/s$ for nonrotating
  SMS solutions.  Crosses (red online) denote the numerical results
  using the exact EOS from Section \ref{sec:nr}. The solid and dashed
  lines (blue online) represent the analytical, leading-order
  perturbative predictions (\ref{deltaMmodel1}) and
  (\ref{deltaMmodel2}) derived using the energy functional method with
  Approximations I and II to the EOS, respectively.  The open circles
  (outlined in black online) and filled squares (green online) denote
  the numerical results from Section \ref{sec:nr}. The triangles
  (purple online) labeled SUS represent numerical results of
  \citet{ShiUS16}, who adopted Approximation II.  The relative change
  in the critical mass increases in magnitude as $\beta$ increases.
  As in Fig.~\ref{xcritfig}, we find that Approximation I is closer to
  the exact treatment of the EOS than Approximation II.  Compare with
  Fig.~2 of Paper II.}
\label{deltaMfig}
\end{figure}

For a SMS supported by radiation pressure alone,
i.e.~pure $n=3$ polytropes, we have $\rho_{0,c}=0$ and hence $x_{\rm
  crit} = 0$, indicating that all of these stars are unstable in relativistic
gravity.  The maximum mass is then given by the Newtonian value
$\bar{M}_{\rm crit}=\bar{M}_{0}^{{\rm sph}} = 4.555$.  In the presence
of gas pressure, the mass will take a maximum at some finite central
density $\rho_{0,c} > 0$, thereby stabilizing those configurations
with central densities smaller than this critical value.  In the
following we parametrize the critical configurations for our EOSs by
the values of $x_{\rm crit}$ and by the relative mass differences
$\delta_{M}^{{\rm sph}}$, defined through
\begin{equation}
\label{defdeltaM}
\bar{M}_{{\rm crit}} = \bar{M}_{0}^{{\rm sph}}\left(1+\delta_{M}^{{\rm sph}}\right).
\end{equation}
In Figs.~\ref{xcritfig} and \ref{deltaMfig} we show results for $x_{\rm crit}$ and $\delta_{M}^{{\rm sph}}$ as a function of $\beta=8k_{B}/s$.

Both Fig.~\ref{xcritfig} and Fig.~\ref{deltaMfig} also include
perturbative results, labeled ``Pert.", that are computed from
analytical, leading-order perturbative expressions derived from a
simple energy functional approach (see Paper II for details).  Both
Approximation I and II lead to identical expressions for $x_{\rm
  crit}$,
\begin{equation}
\label{xcritmodel1}
x_{{\rm crit}} = \frac{k_{2}}{4k_{4}}\beta
\end{equation}
(see eqs.~(49) and (56) in Paper II, hereafter (II.49) and (II.56)),
where $k_{2}=0.63899$ \citep{LaiRS93} and $k_{4}=0.918294$
\citep{ShaT83}.  The two approximations differ, however, in their
predictions for the corrections to the mass.  For Approximation I,
this correction is
\begin{equation}
\label{deltaMmodel1}
\delta_{M}^{{\rm sph,I}} = \left(\frac{3}{4}\ln\frac{k_{2}}{4k_{4}}+2\ln\beta+\frac{3}{2}C\right)\beta,
\end{equation}
(see (II.51)) with
\begin{equation}
\label{defC}
C = \frac{k_{\tau}}{2} - \frac{1}{3}\ln\bar{M}_0^{\rm sph} - \frac{1}{6}\ln\eta - \frac{1}{3},
\end{equation}
$k_{\tau} = -0.45928$, and $\eta$ given by (\ref{etaApp1def}), while
for Approximation II it is
\begin{equation}
\label{deltaMmodel2}
\delta_{M}^{{\rm sph,II}} = \left(\frac{3}{4}\ln\frac{k_{2}}{4k_{4}}+\frac{3}{4}\ln\beta-\frac{1}{2}\ln\bar{M}_{0}^{{\rm sph}}\right)\beta
\end{equation}
(see (II.60)).

Fig.~\ref{xcritfig} (compare with Fig.~1 of Paper II) shows that when
a SMS is partially supported by gas pressure ($\beta>0$) it is
stabilized against collapse ($x_{\rm crit}>0$) for central densities
below $\rho_{0,c}$.  The numerical solution using the exact EOS falls
between the solutions using Approximations I and II.  As suggested in
Paper II, Approximation I is closer to the exact solution, despite
Approximation II agreeing better with the perturbative prediction.

Fig.~\ref{deltaMfig} (compare with Fig.~2 of Paper II) shows that the
relative change in the critical mass increases in magnitude as $\beta$
increases.  As in Fig.~\ref{xcritfig}, the numerical solution from
handling the EOS exactly falls between the solutions using
Approximations I and II, but is much closer to the results of
Approximation I.  Also included in this plot are numerical results of
\citet{ShiUS16}, labeled SUS, who adopted Approximation II.  Not
surprisingly, their results agree very well with our corresponding
results.

%
\section{Rotating Supermassive Stars}
\label{sec:rot}
%

As discussed in Paper I,
rotation can stabilize a SMS even when it is supported by a pure
radiation gas, i.e.~an $n=3$ polytrope.  In fact, for maximally
rotating SMS, i.e.~stars rotating uniformly at the mass-shedding limit, the
critical configuration marking the onset of a radial instability is
characterized by unique values of the dimensionless parameters
\begin{align}
\label{x_0}
x_0 & \simeq 5.97 \times 10^{-3}, \\
\label{j_0}
j_0 & \simeq 0.919,
\end{align}
where $j = J / M^2$ is the dimensionless angular momentum, and
\begin{equation}
\label{M_0}
\bar M_0 \simeq 4.56
\end{equation}
(see Section \ref{sec:rot:radiation} below).  In this Section we
evaluate how changes in these parameters due to the presence of gas
pressure are affected by the different treatments of the gas pressure.
Specifically, we will compute changes $\delta_x$, $\delta_j$ and
$\delta_M$, defined by
\begin{align}
\label{xrot}
x & = x_{0}\left(1+\delta_{x}\right), \\
\label{jrot}
j & = j_{0}\left(1+\delta_{j}\right), \\
\label{Mrot}
\bar{M} & = \bar{M}_{0}\left(1+\delta_{M}\right),
\end{align}
using the exact and approximate treatments of the gas pressure.  As in
Section \ref{sec:nr} we will also compare these changes with the
perturbative expressions of Paper II.


\subsection{Numerical Method}
\label{sec:rot:nummeth}

We use a version of the RNS code (see \cite{SteF95}) slightly modified
for use with SMSs.  We use the tabulated EOS option with the EOSs
discussed in Section \ref{sec:eos} and tables assembled using code
discussed in Section \ref{sec:eos:numerical}.  We change the default
surface values for energy density, pressure, and enthalpy in the
example {\tt main.c} to zero for tabulated EOSs.  We also make a
radial step size in the RNS code's TOV-solver in {\tt equil.c} six
orders of magnitude larger.  Both changes are needed because SMSs are
far less dense than neutron stars, and much larger.  We add a
high-resolution grid option to the makefile for these calculations,
increasing the number of angular gridpoints to 801 and the number of
radial gridpoints to 1601.  Given an EOS and a central energy density
the example RNS code spins up a TOV solution until the star reaches
mass-shedding, finding many intermediate configurations along the way.
For a given EOS we consider many different central densities, allowing
us to compute the data displayed in Figs.~\ref{polytropefig} and
\ref{bigcomparisonfig}.

The curves of constant $\bar{J}$ in Figs.~\ref{polytropefig} and
\ref{bigcomparisonfig} are constructed by interpolation.  Stable and
unstable configurations are separated by locating the maximum mass
$\bar{M}$ along curves of constant $\bar{J}$ (see \citet{FriIS88} and discussion in \citet{BauS10}).  We mark these turning
points in Figs.~\ref{polytropefig} and \ref{bigcomparisonfig} with
black dots.  The turning point corresponding to the maximally rotating
configuration is marked separately as the critical point.

\subsection{Pure Radiation Fluid}
\label{sec:rot:radiation}

\begin{figure}
\begin{center}
\includegraphics[width=0.47 \textwidth]{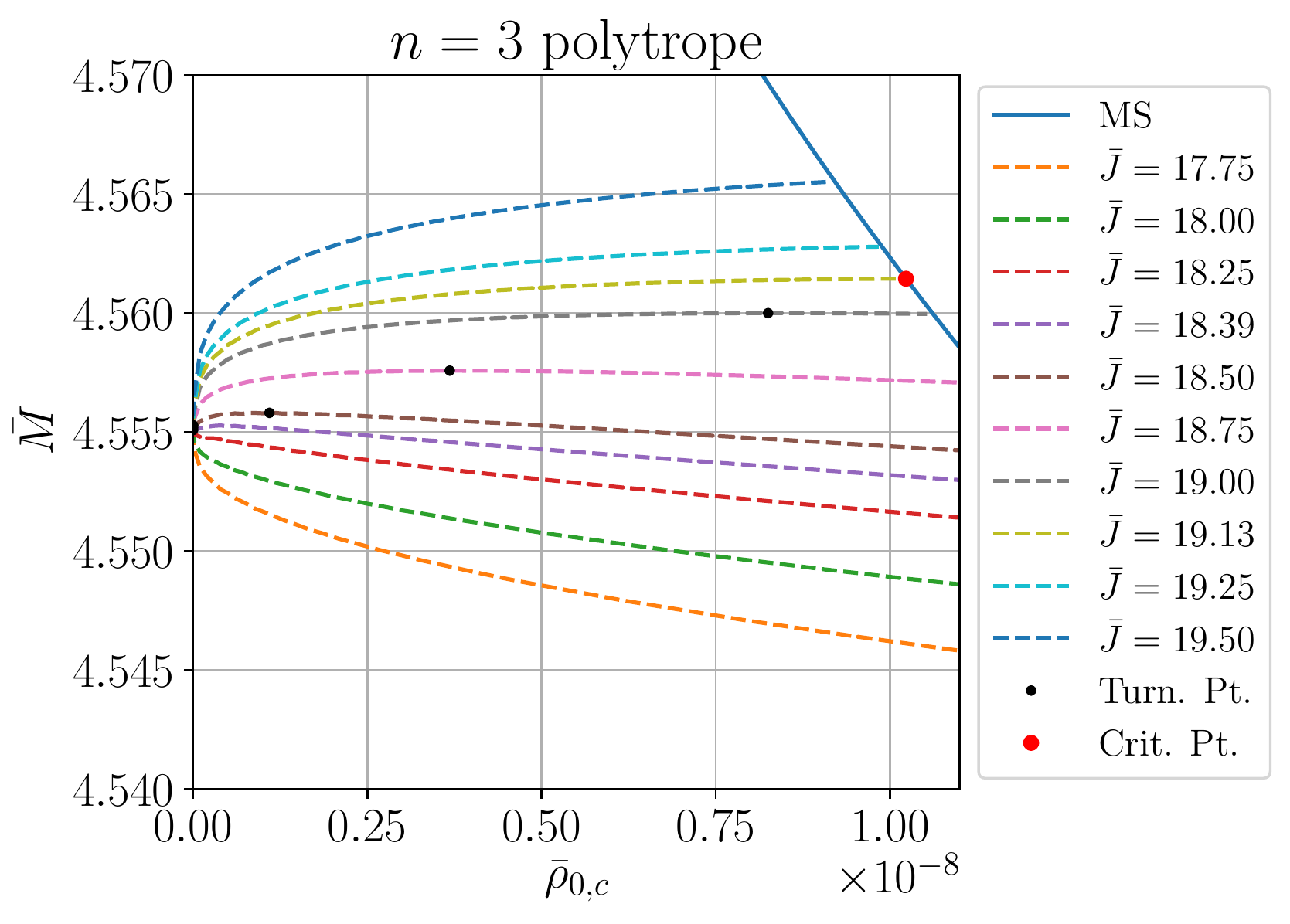}
\end{center}
\caption{The mass-shedding curve (MS in the legend; solid blue online)
  and curves of constant $\bar{J}$ for a SMS supported by a pure
  radiation fluid, i.e.~an $n=3$ polytrope.  Black dots denote the
  turning points (``Turn. Pt." in the legend), i.e.~maxima of curves
  of constant $\bar{J}$.  The larger dot (red online) denotes the
  critical configuration (``Crit. Pt." in the legend), defined as the
  intersection of the turning-point and mass-shedding curves.
  (Compare Fig.~2 in Paper I.)}
\label{polytropefig}
\end{figure}

We start our analysis for a SMS supported by a pure radiation fluid,
i.e.~for an $n=3$ polytrope, essentially reproducing the numerical
analysis of Paper I.  Our results are shown in
Fig.~\ref{polytropefig}.  In particular, we identify the critical
configuration as the mass-shedding configuration at the onset of
radial instability.  The dimensionless parameters $x_0$, $j_0$ and
$\bar M_0$ characterizing this critical configuration are given in
eqs.~(\ref{x_0}) through (\ref{M_0}) above, see also Table \ref{bigtable}.  
We note that these values
differ slightly from those computed in Paper I; we believe that the
differences are due to the significantly higher numerical resolution
used in our work here.  In the following Sections we evaluate how gas
pressure, treated both exactly and approximately, affects these
critical parameters.

\subsection{Exact Approach}
\label{sec:rot:exact}

\begin{figure*}
\begin{center}
\includegraphics[scale=0.35]{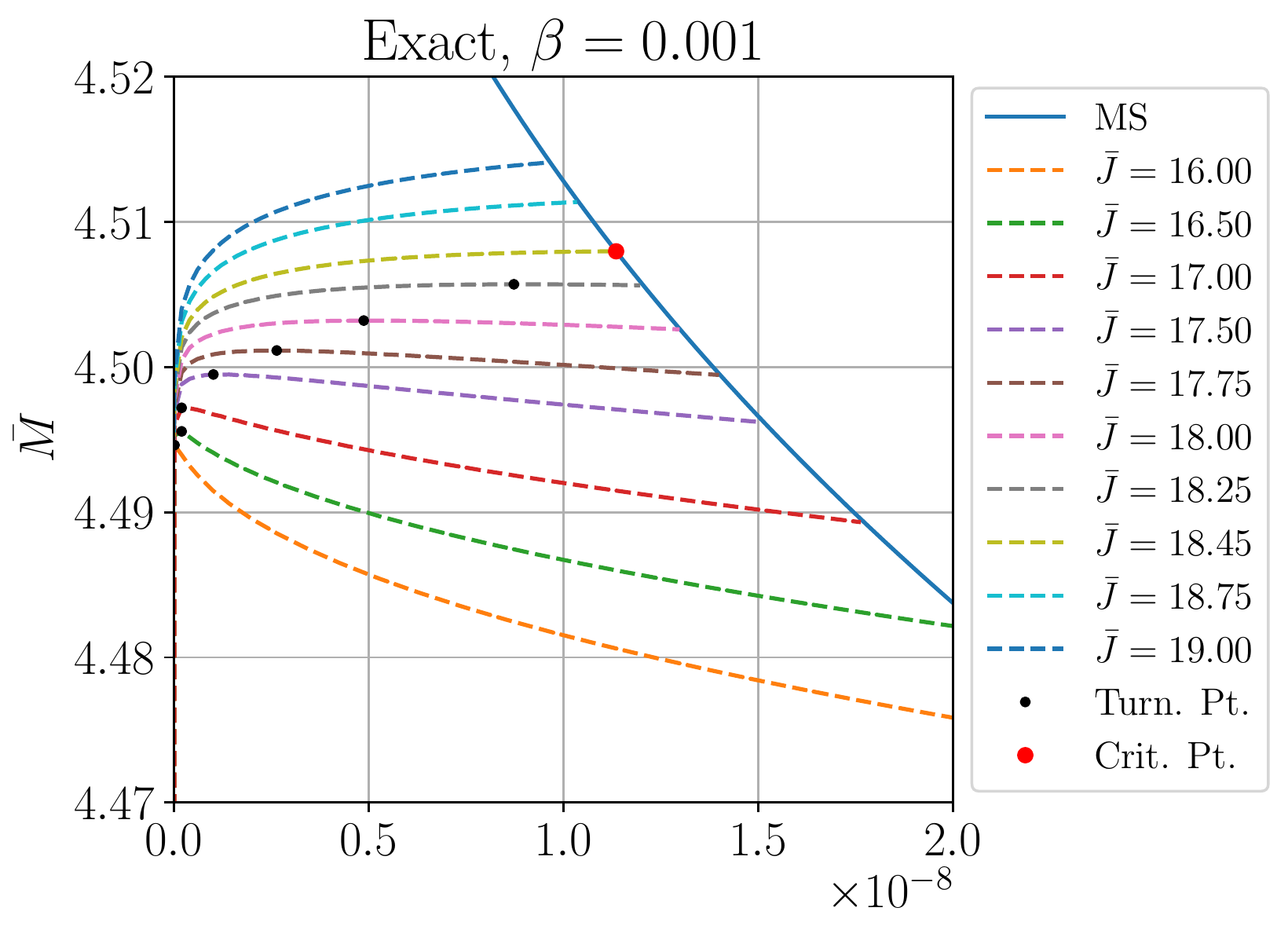}\hspace{0.01\textwidth}\includegraphics[scale=0.35]{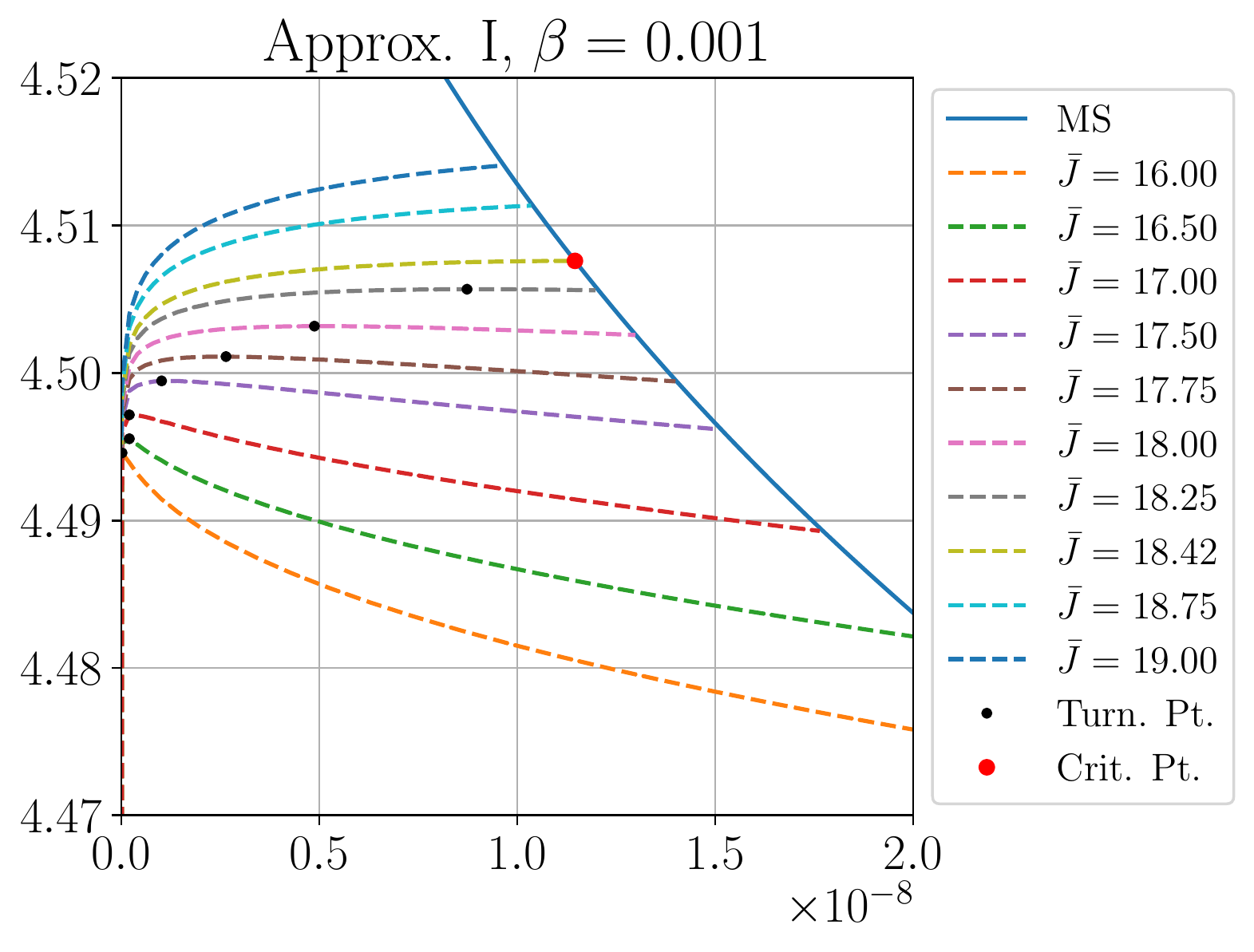}\hspace{0.01\textwidth}\includegraphics[scale=0.35]{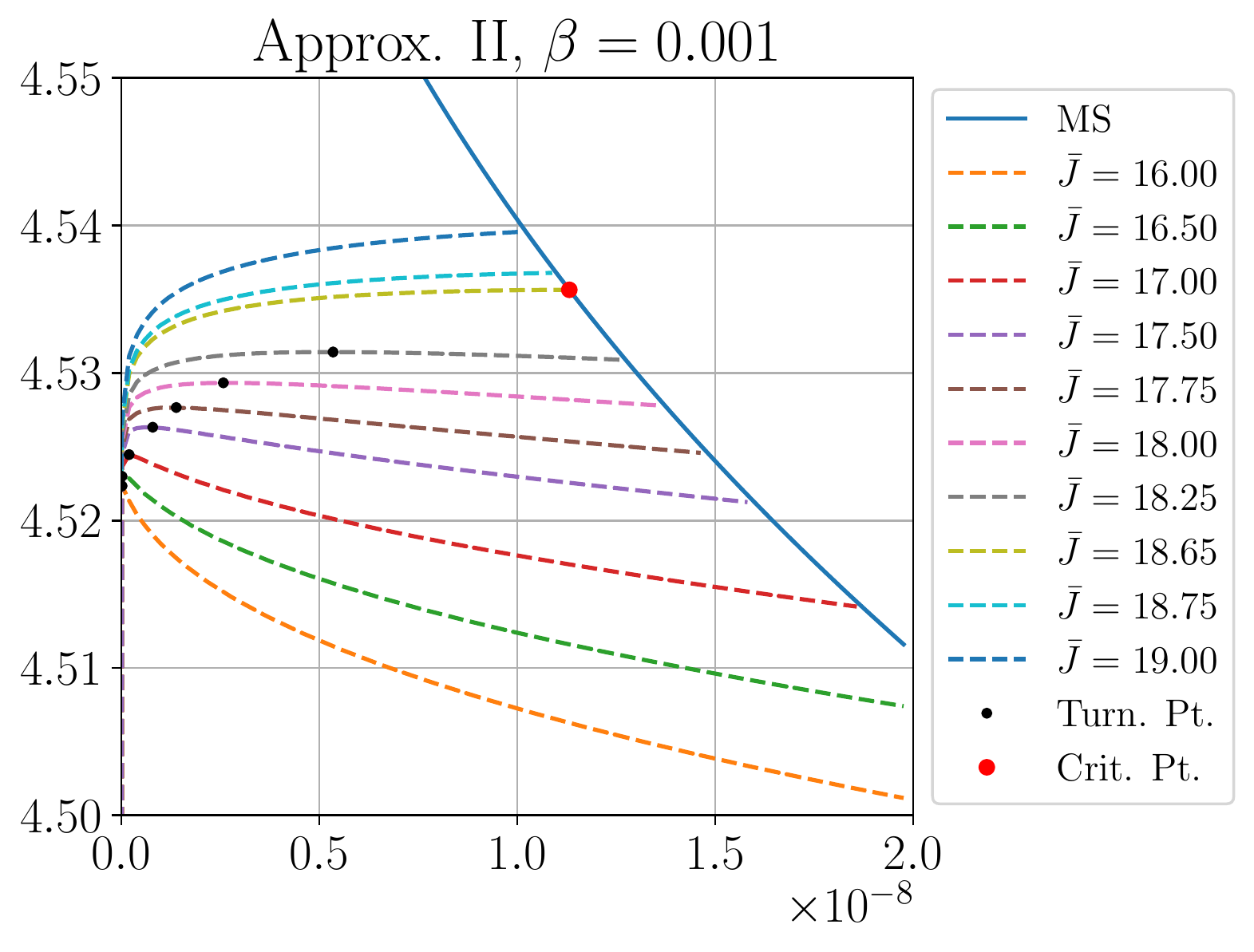}\\
\includegraphics[scale=0.35]{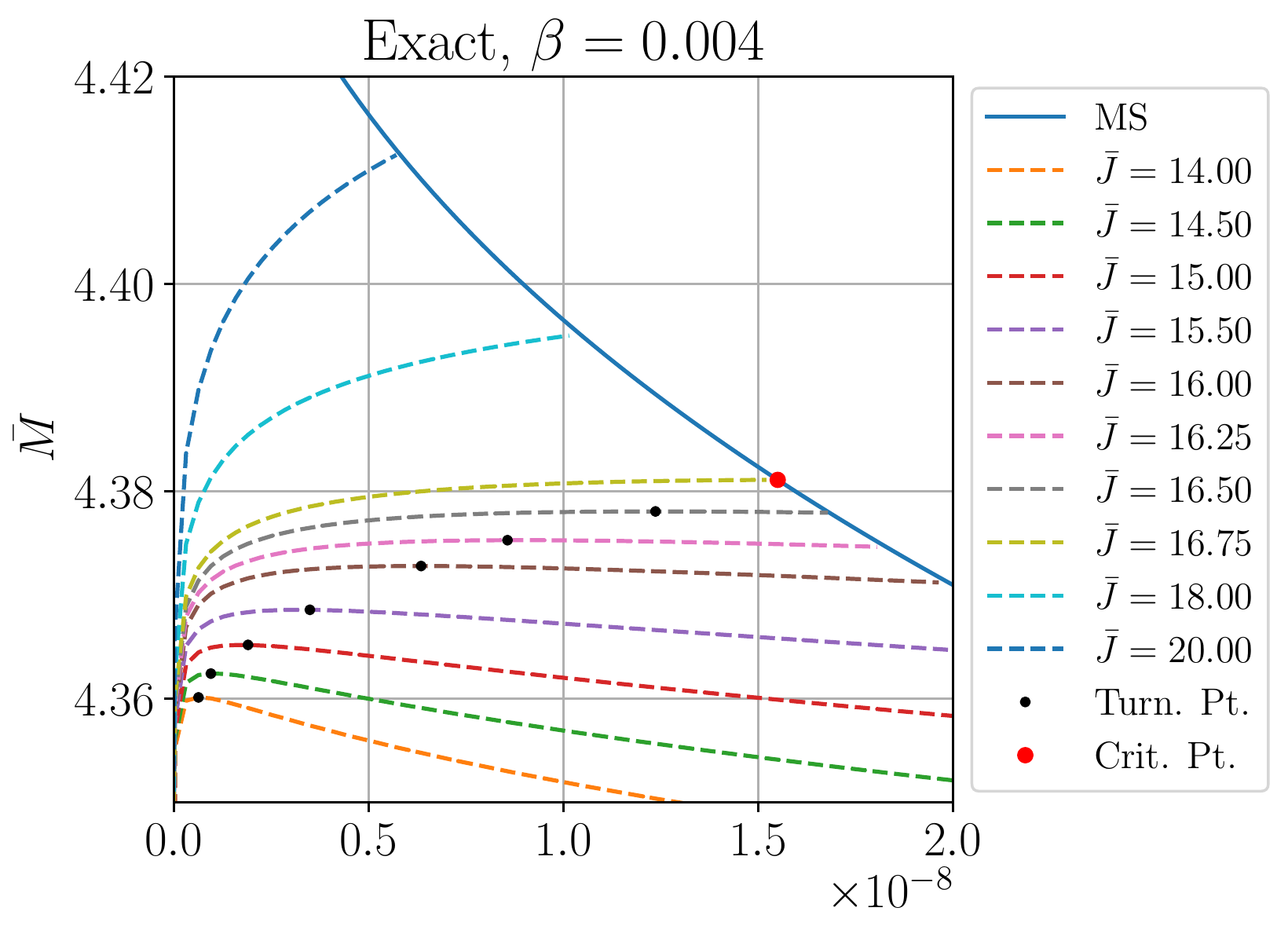}\hspace{0.01\textwidth}\includegraphics[scale=0.35]{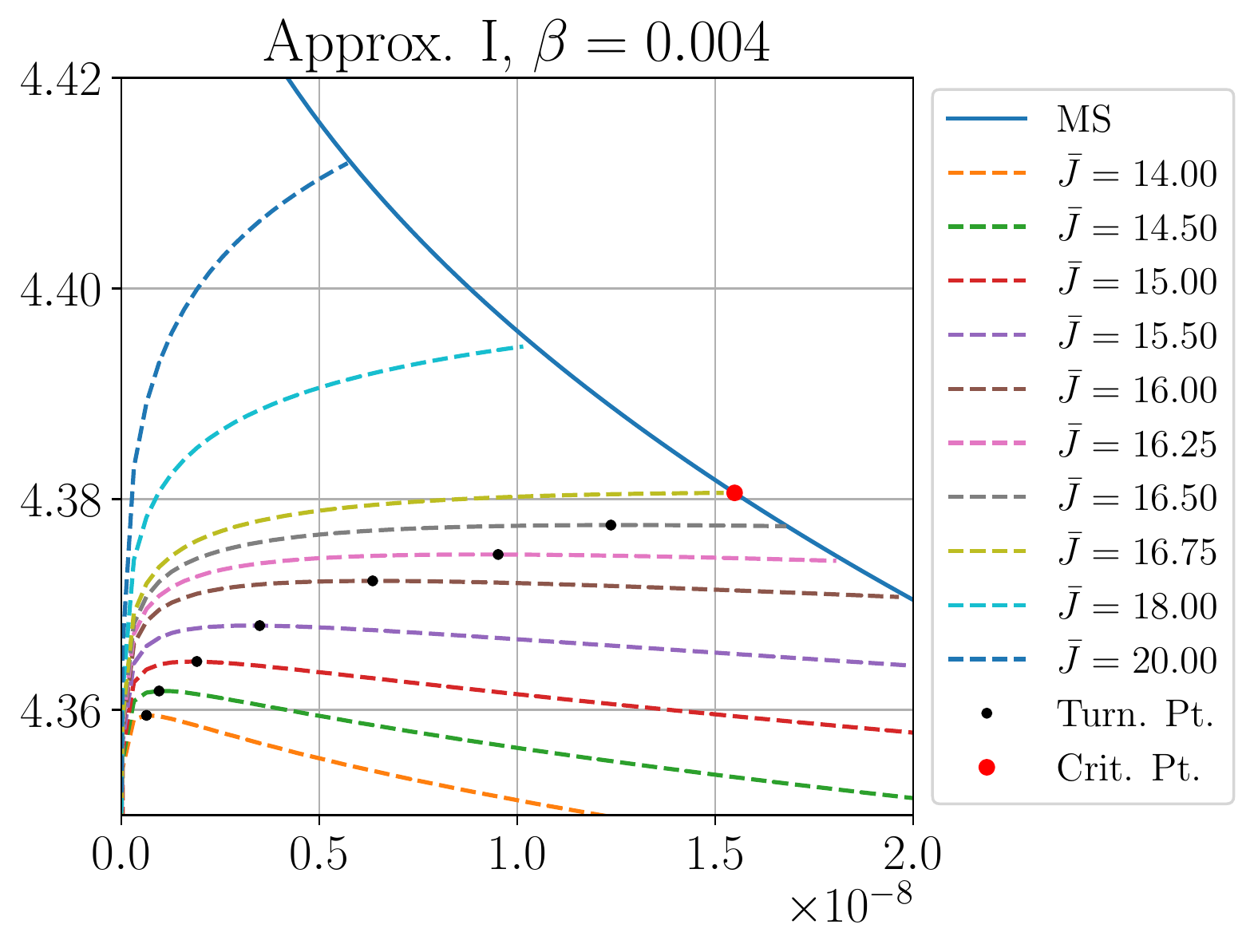}\hspace{0.01\textwidth}\includegraphics[scale=0.35]{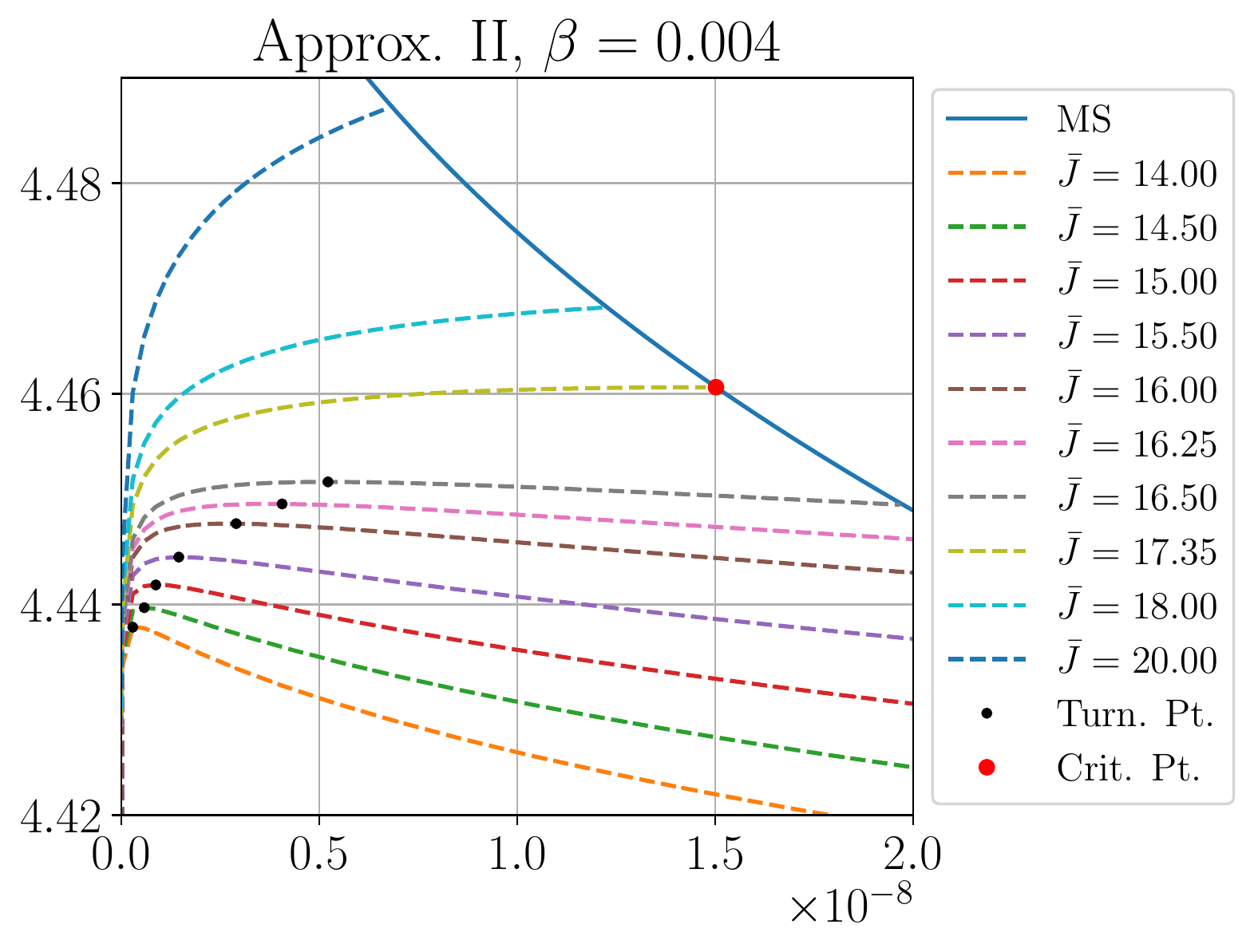}\\
\includegraphics[scale=0.35]{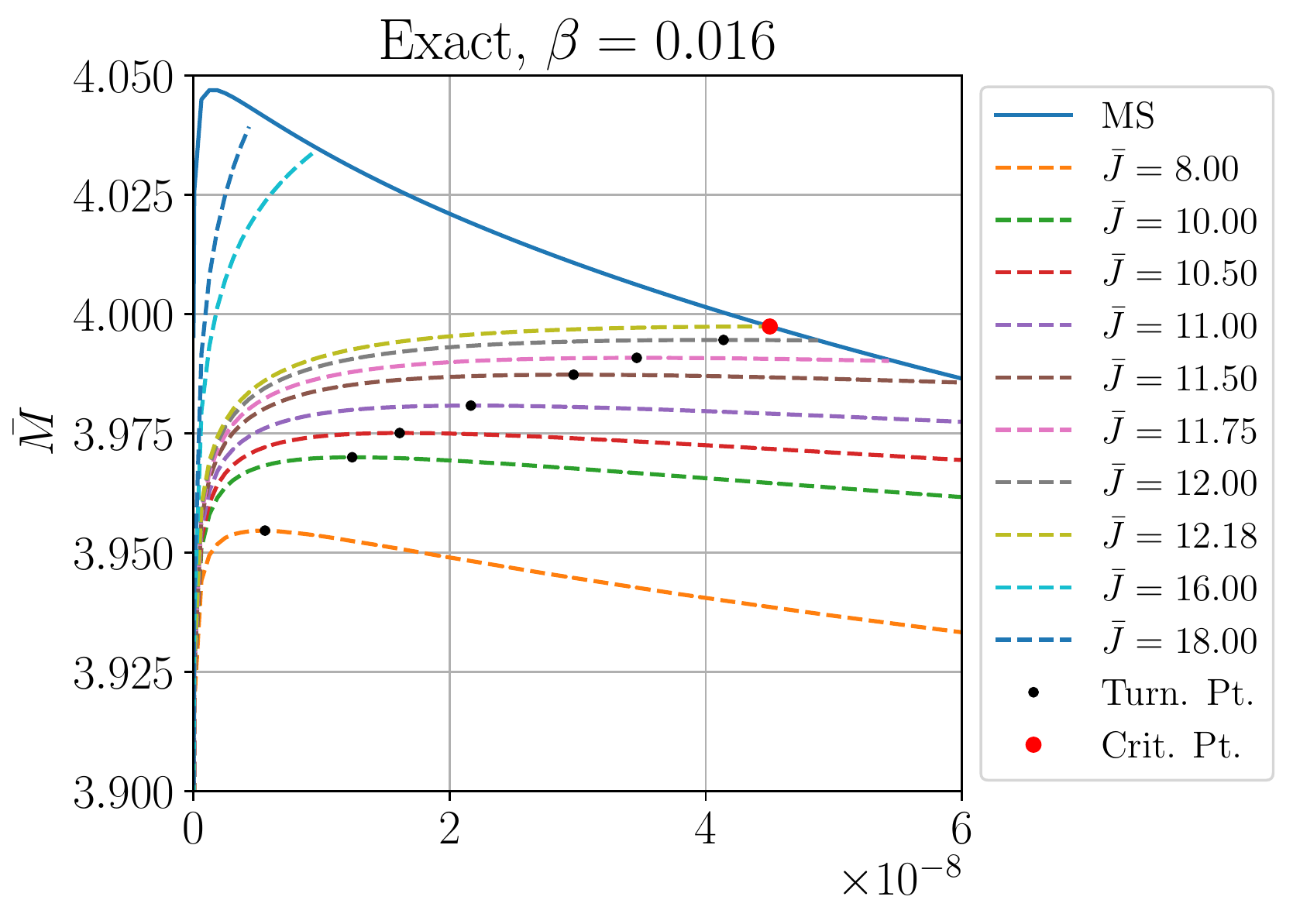}\hspace{0.01\textwidth}\includegraphics[scale=0.35]{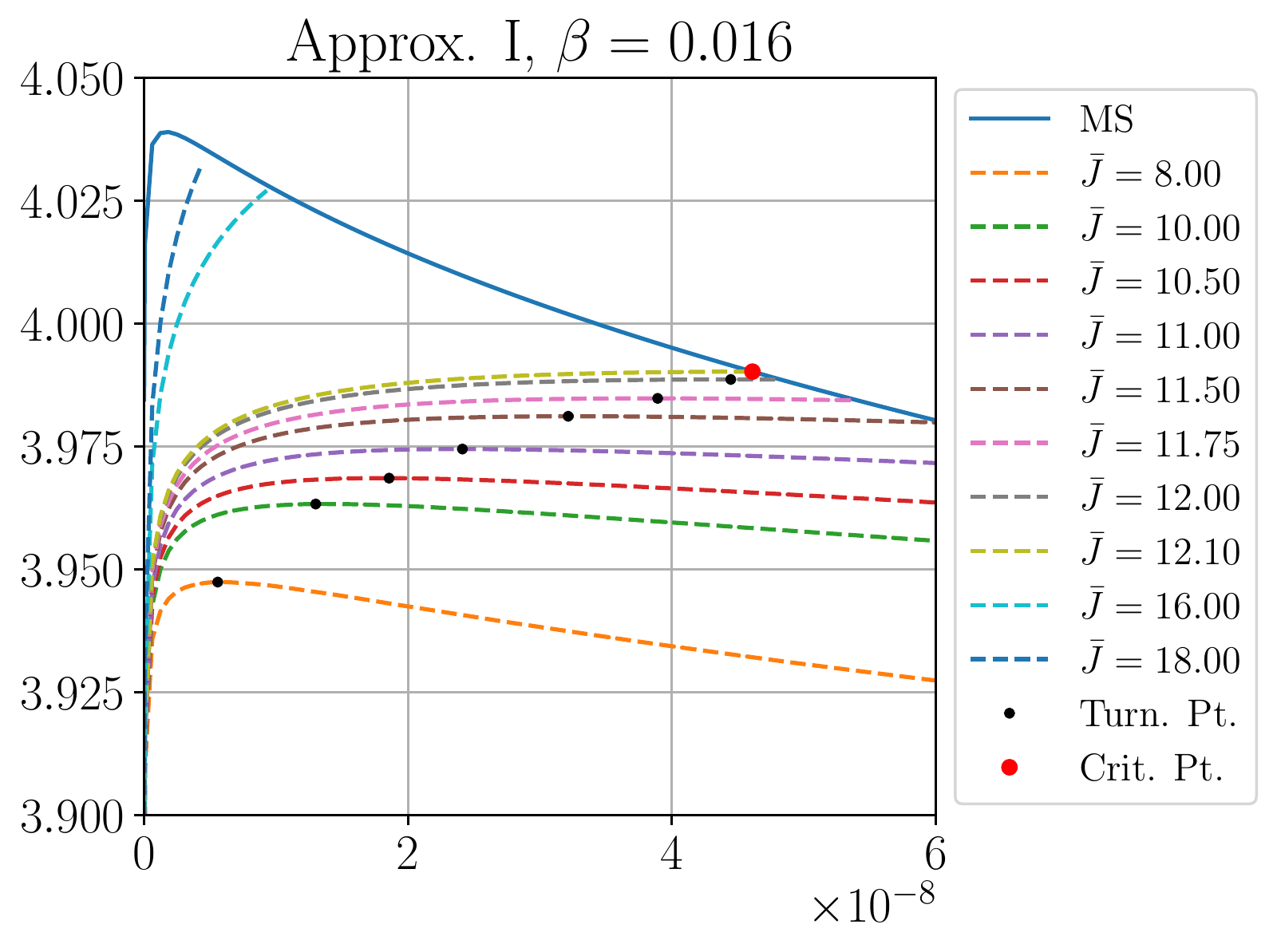}\hspace{0.01\textwidth}\includegraphics[scale=0.35]{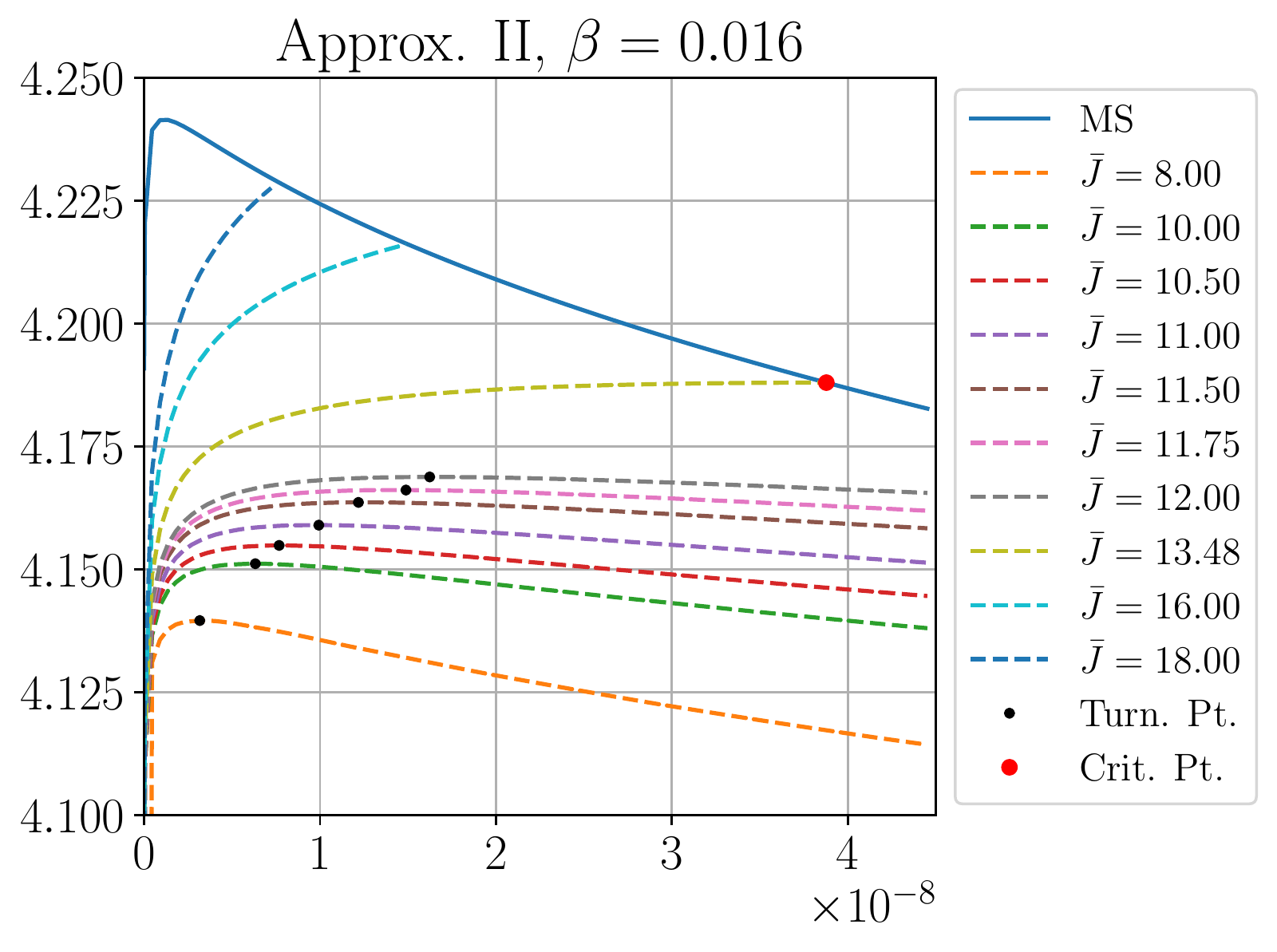}\\
\includegraphics[scale=0.35]{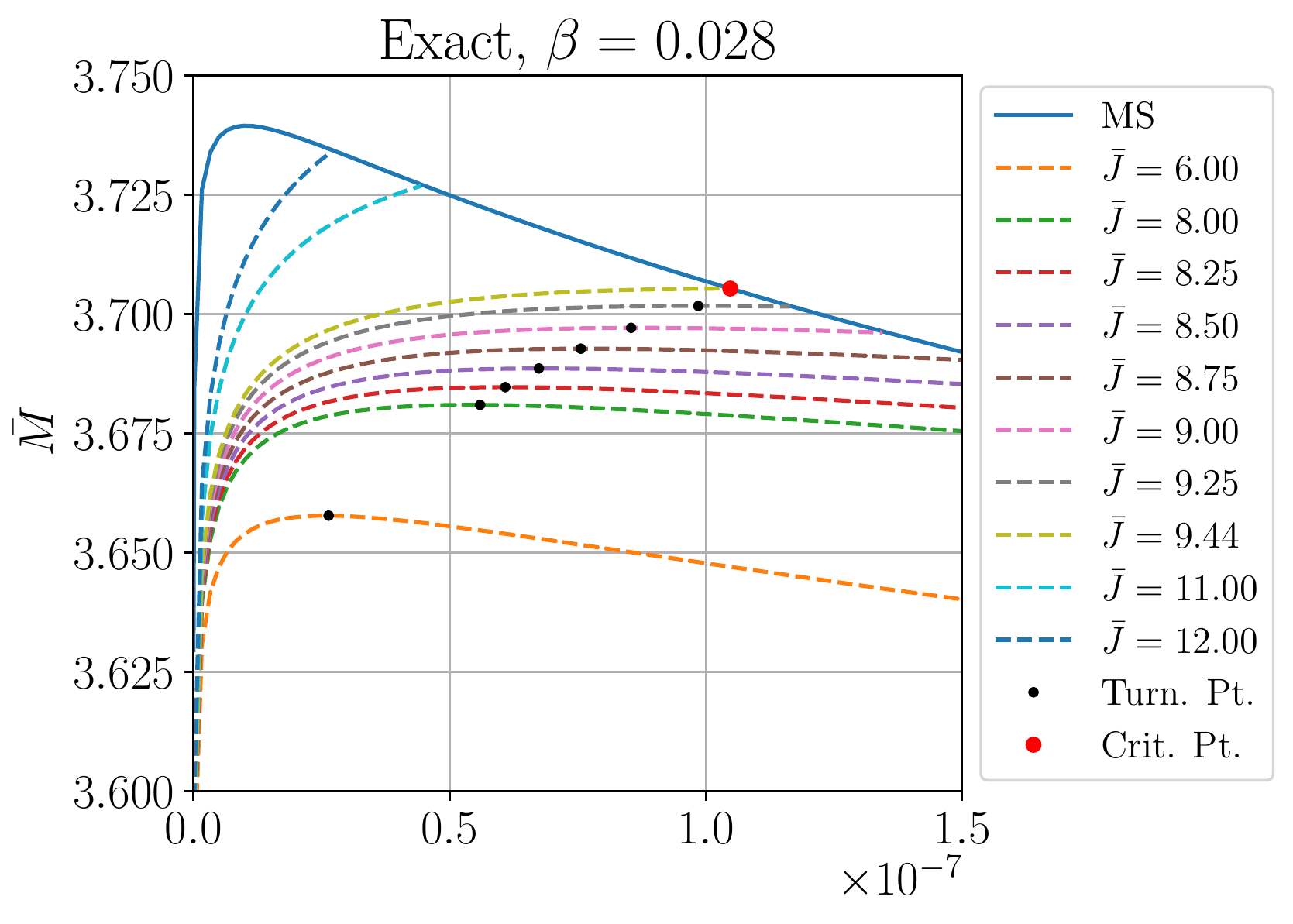}\hspace{0.01\textwidth}\includegraphics[scale=0.35]{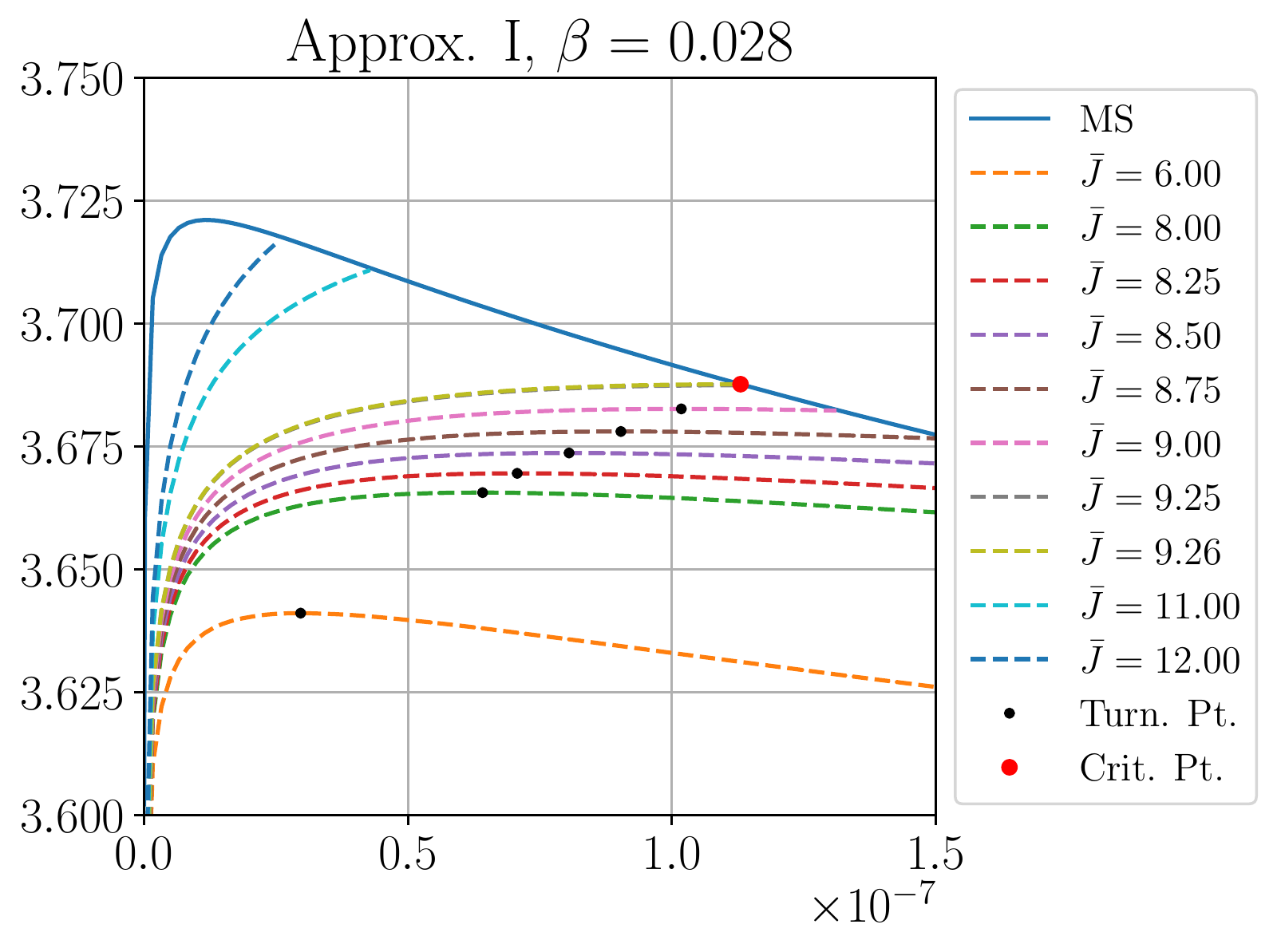}\hspace{0.01\textwidth}\includegraphics[scale=0.35]{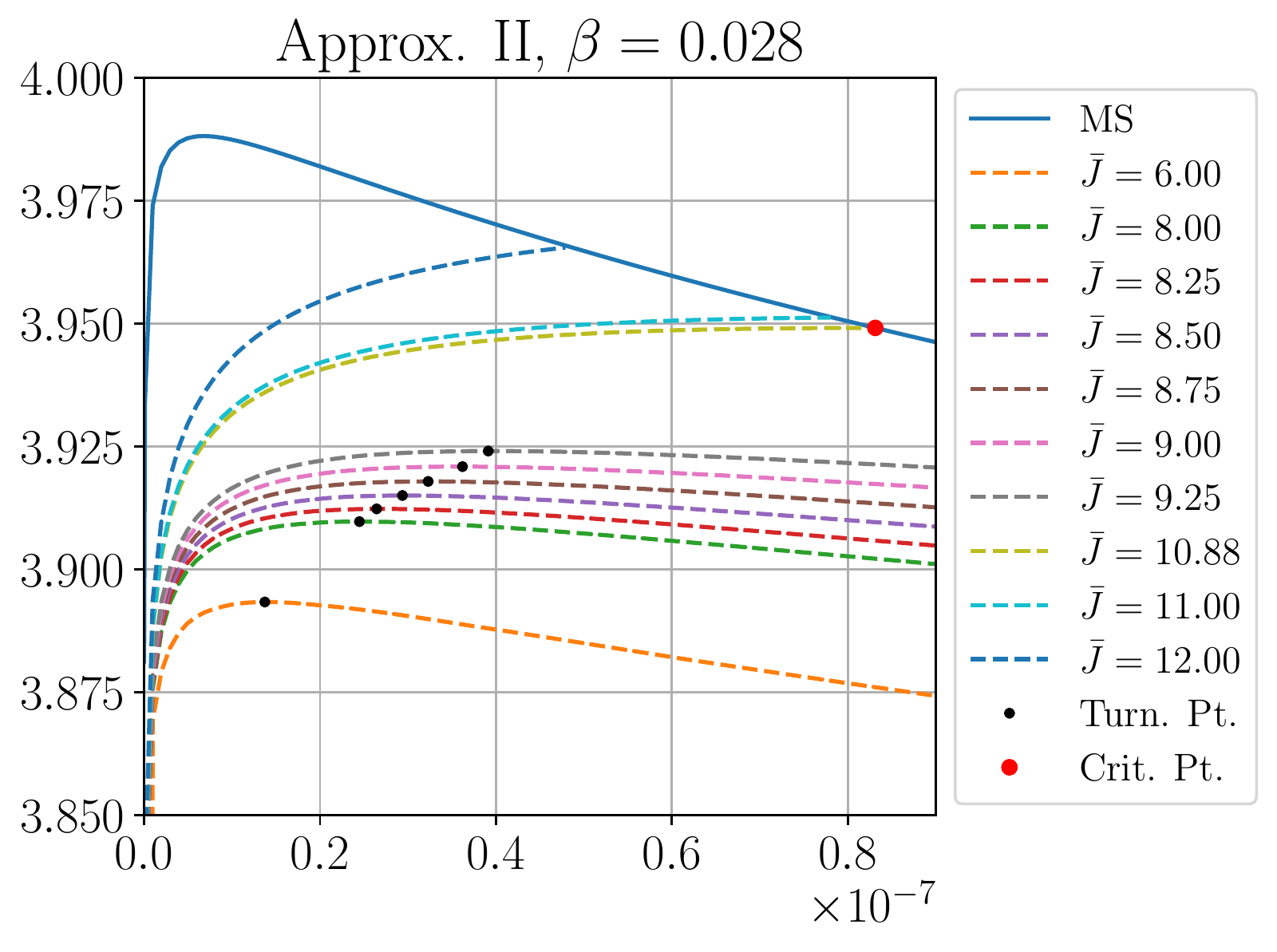}\\
\includegraphics[scale=0.35]{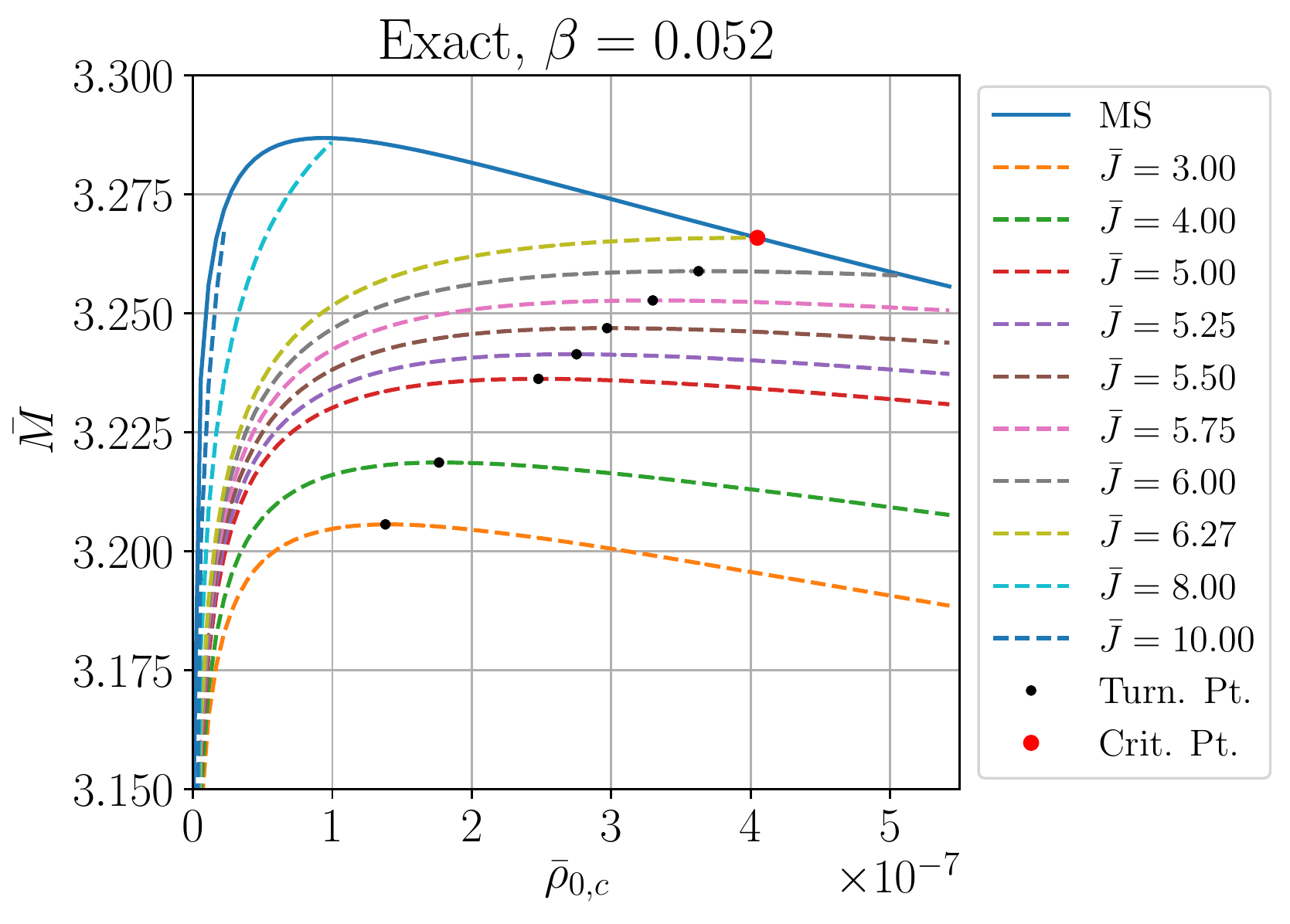}\hspace{0.01\textwidth}\includegraphics[scale=0.35]{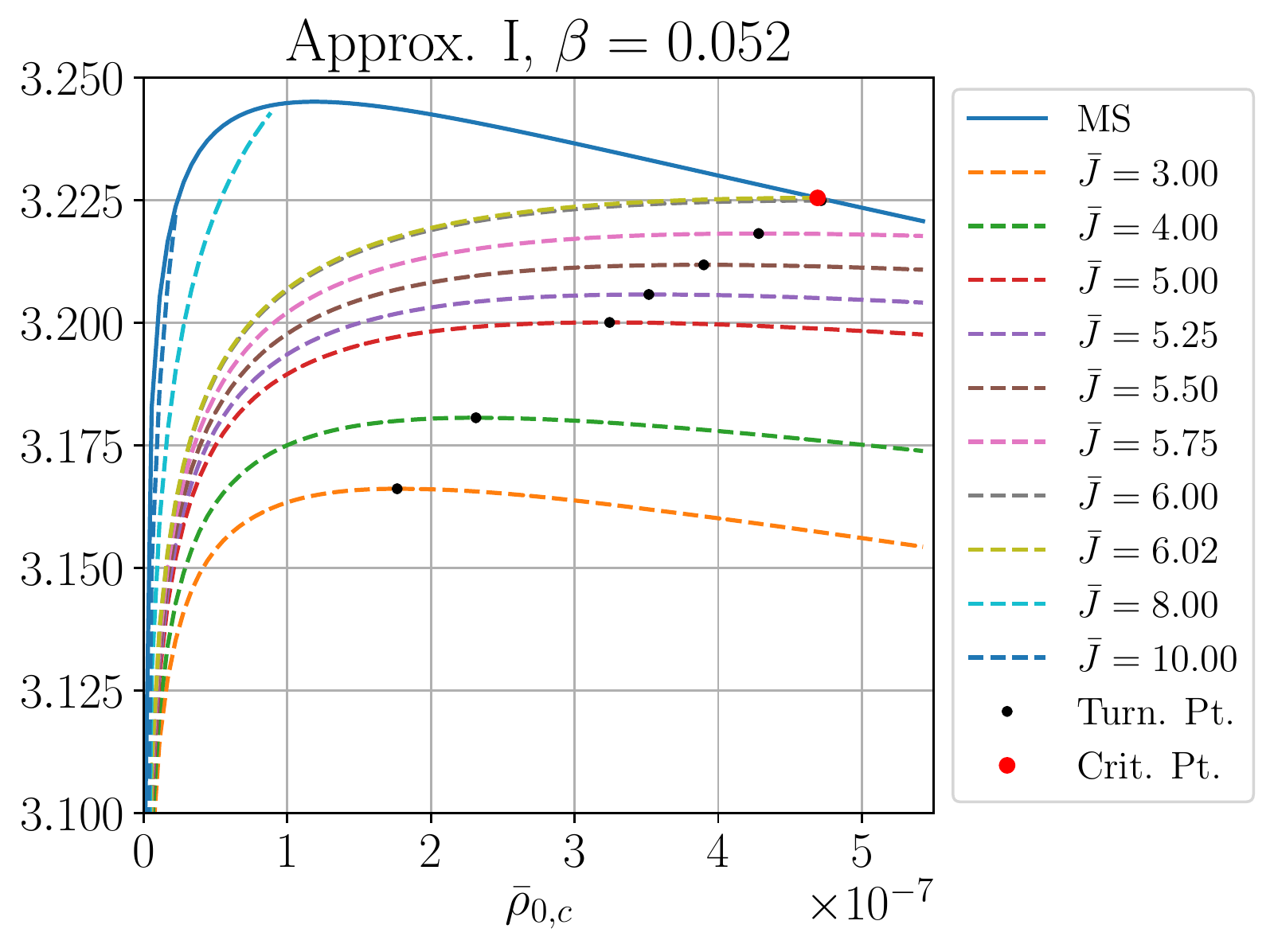}\hspace{0.01\textwidth}\includegraphics[scale=0.35]{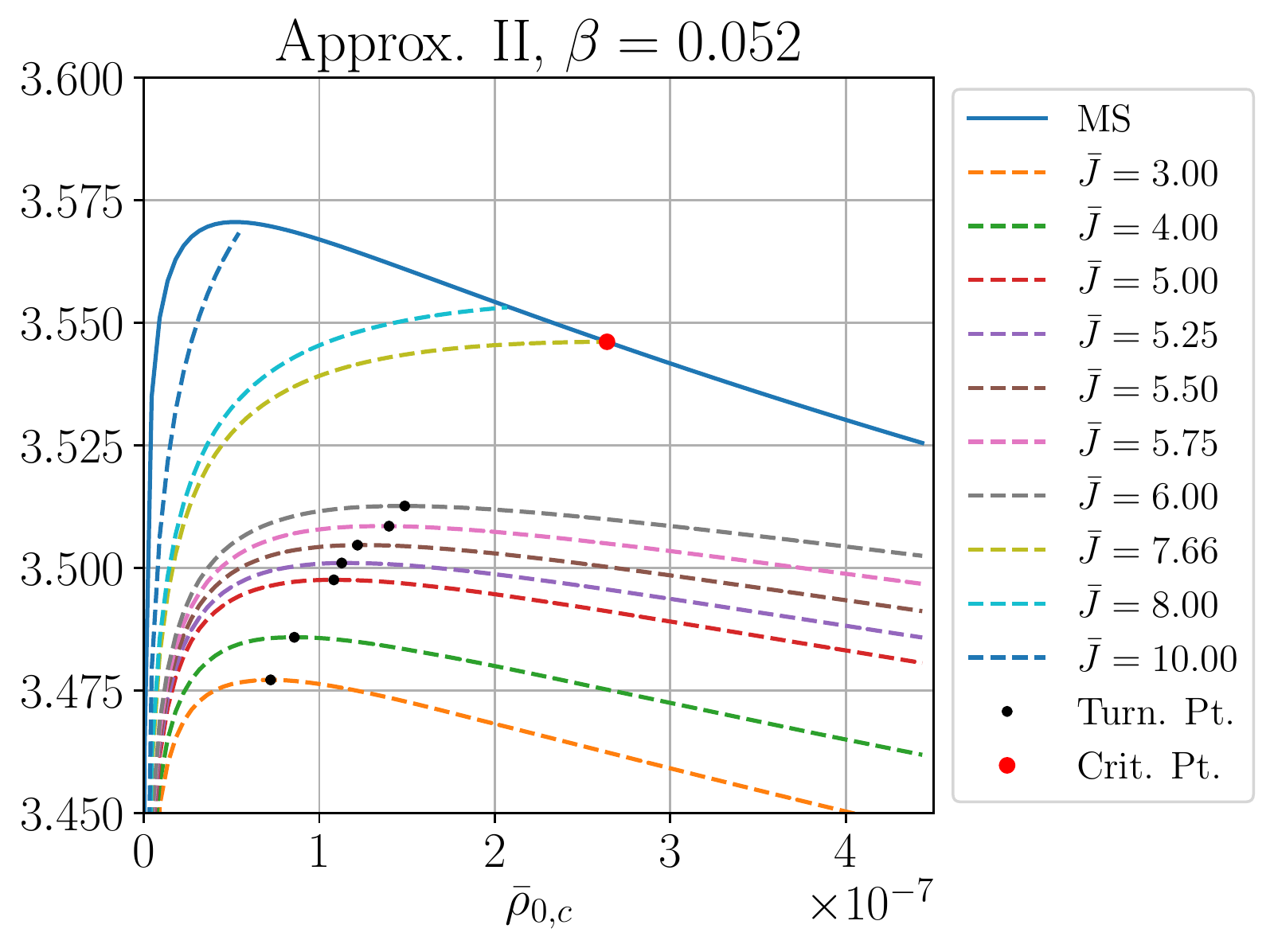}\\
\end{center}
\caption{The mass-shedding curve (MS in each legend; blue online) and
  curves of constant $\bar{J}$ for different treatments of the EOS and
  different values of $\beta$.  The left column shows results for the
  exact treatment of the EOS, the middle column for Approximation I,
  and the right column for Approximation II.  Black dots mark turning
  points, and red dots the critical configurations, as in
  Fig.~\ref{polytropefig}.}
\label{bigcomparisonfig}
\end{figure*}

\begin{figure*}
\begin{center}
\includegraphics[width=0.47 \textwidth]{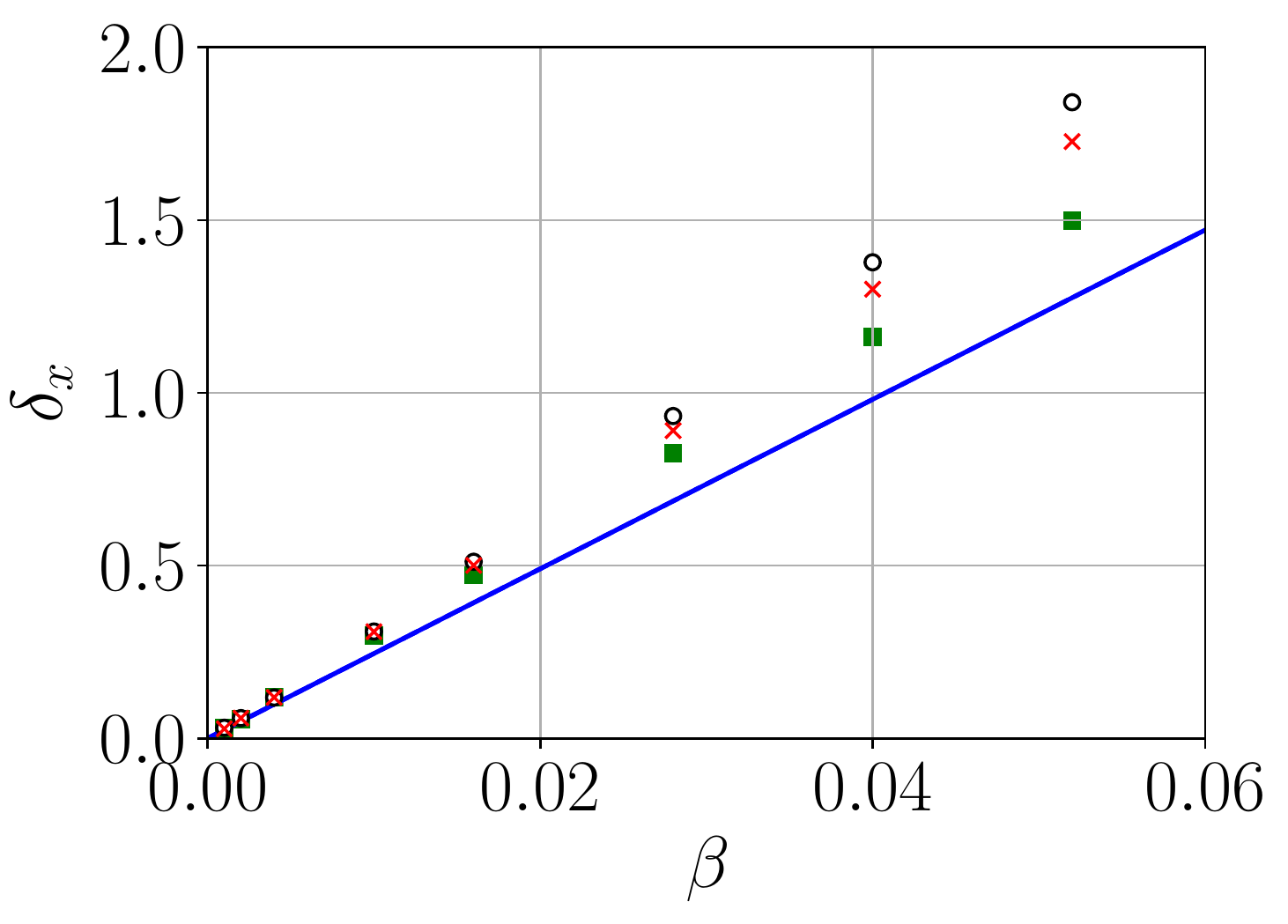}~~~
\includegraphics[width=0.47 \textwidth]{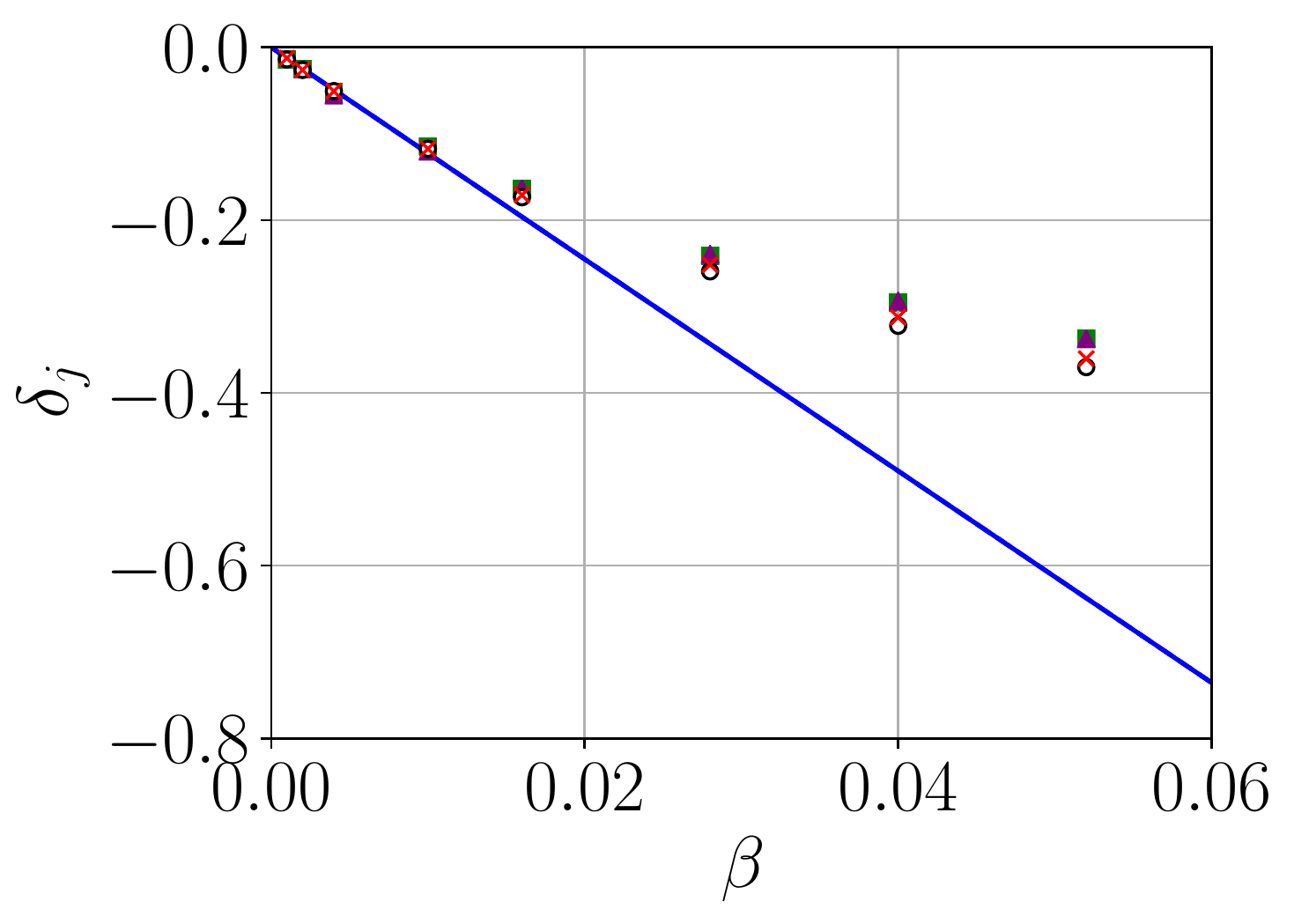}

\includegraphics[width=0.47 \textwidth]{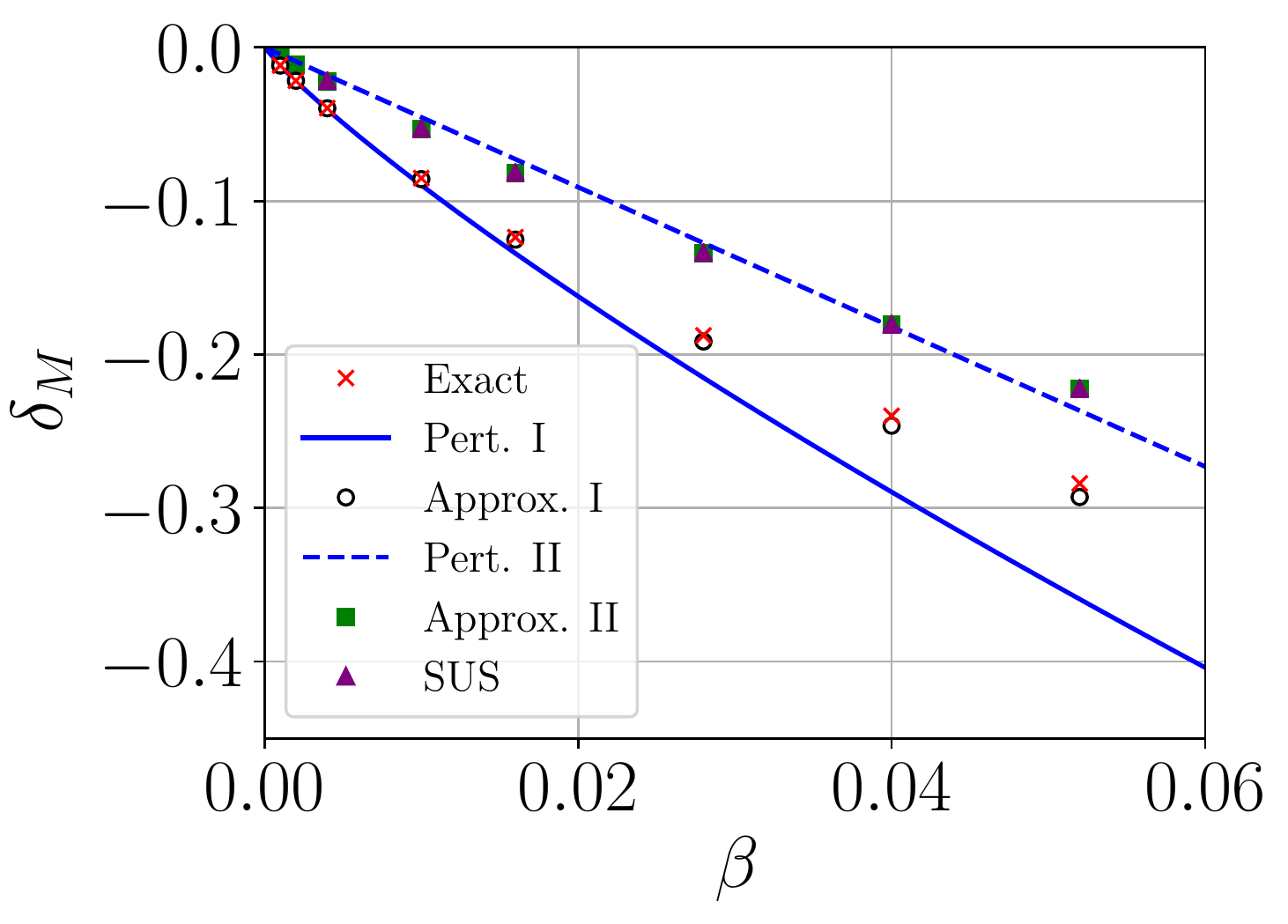}
\end{center}
\caption{Effects of gas pressure on the dimensionless parameter $x$
  (top left panel), the dimensionless angular momentum $j$ (top right panel),
  and the rescaled mass $\bar{M}$ (bottom panel) of the critical
  configuration of rotating SMSs.  We show $\delta_{x}$, $\delta_{j}$,
  and $\delta_{M}$, defined in eqs.~(\ref{xrot}) -- (\ref{Mrot}), as a
  function of $\beta$.  Crosses (red online) denote numerical results
  from the exact treatment of the EOS.  The solid and dashed lines
  (blue online) represent the analytical, leading-order perturbative
  expressions from applying the energy functional approach to
  Approximation I and Approximation II.  The solid and dashed lines
  coincide for $\delta_{x}$ and $\delta_{j}$.  The open circles
  (outlined in black online) and filled squares (green online) denote
  our corresponding numerical results for Approximations I and II.
  The triangles (purple online) in the $\delta_{j}$ and $\delta_{M}$
  plots represent the numerical results of \citet{ShiUS16}, who
  adopted Approximation II; they agree so well with our results for
  Approximation II that they are difficult to distinguish in the plot.
  We find that Approximation I is closer to the exact treatment of the
  EOS than Approximation II.  Compare with Fig.~3 of Paper II.}
\label{comparetofig3paperII}
\end{figure*}

\begin{table*}
\begin{tabular}{c|c|c|c|c|c|c}
$\beta$&	$R_{p}/M$&	$J/M^{2}$&	$\bar{M}$&	$M~[M_{\sun}]$&	$\rho_{c}~[{\rm g/cm^{3}}]$&	$T~[K]$\\
\hline 
$0$     &	$380$&		$0.919$&	$4.56$&		-- &			-- &				--\\
$0.001$&	$369$&		$0.908$&	$4.51$&		$7.17\times 10^7$&			$2.78\times 10^{-5}$&		$1.22\times 10^{7}$\\
$0.002$&	$358$&		$0.895$&	$4.46$&		$1.77\times 10^7$&			$4.96\times 10^{-4}$&		$2.53\times 10^{7}$\\
$0.004$&	$338$&		$0.873$&	$4.38$&		$4.35\times 10^6$&			$9.72\times 10^{-3}$&		$5.39\times 10^{7}$\\
$0.010$&	$288$&		$0.811$&	$4.17$&		$6.64\times 10^{5}$&	$0.670$&			$1.61\times 10^{8}$\\
$0.016$&	$250$&		$0.762$&	$4.00$&		$2.48\times 10^{5}$&	$7.22$&				$3.02\times 10^{8}$\\
$0.028$&	$196$&		$0.688$&	$3.71$&		$7.51\times 10^{4}$&	$158$&				$6.92\times 10^{8}$\\
$0.040$&	$159$&		$0.632$&	$3.47$&		$3.45\times 10^{4}$&	$1.35\times 10^{3}$&		$1.24\times 10^{9}$\\
$0.052$&	$133$&		$0.588$&	$3.27$&		$1.92\times 10^{4}$&	$7.25\times 10^{3}$&		$1.97\times 10^{9}$\\
\end{tabular}
\caption{Critical configuration parameters for an $n=3$ polytrope (first row) 
and for the exact treatment of the EOS (other rows).  Effects of electron-positron
pair production become important for temperatures greater than about $10^9$ K \protect\citep[see, e.g.,][]{KipWW12},
but are ignored in our treatment here.}
\label{bigtable}
\end{table*}

We start with the exact treatment of the EOS, as described in Section
\ref{sec:eos:exact}.  To do so, we run the RNS code with the
corresponding tabulated EOS for different values of $\beta = 8 k_B /
s$.  For each value of $\beta$ we again choose a number of different
values for the central density, and let the RNS code spin the star up
to mass shedding.  Results from these calculations are shown in the
left column of Fig.~\ref{bigcomparisonfig}.  We determine the critical
configurations, marked by the red dots in Fig.~\ref{bigcomparisonfig},
as before, and compute their physical parameters (see Table \ref{bigtable}).  
Finally, we compute
the corresponding changes from eqs.~(\ref{xrot}) ---(\ref{Mrot}), and
plot these changes in Fig.~\ref{comparetofig3paperII}.

\begin{figure*}
\begin{center}
$\vcenter{\hbox{\includegraphics[width=0.47 \textwidth]{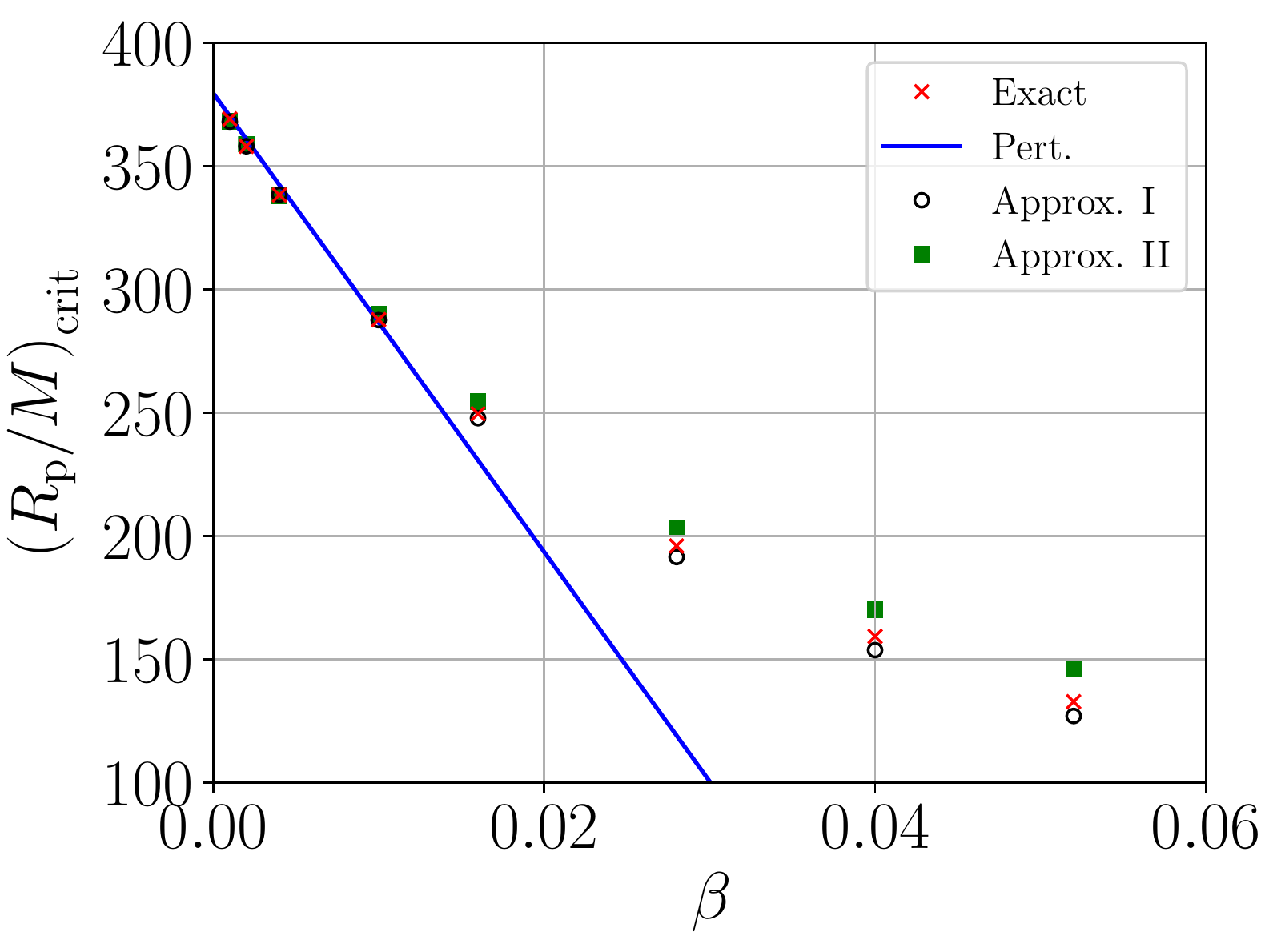}}}$~~~
$\vcenter{\hbox{\includegraphics[width=0.47 \textwidth]{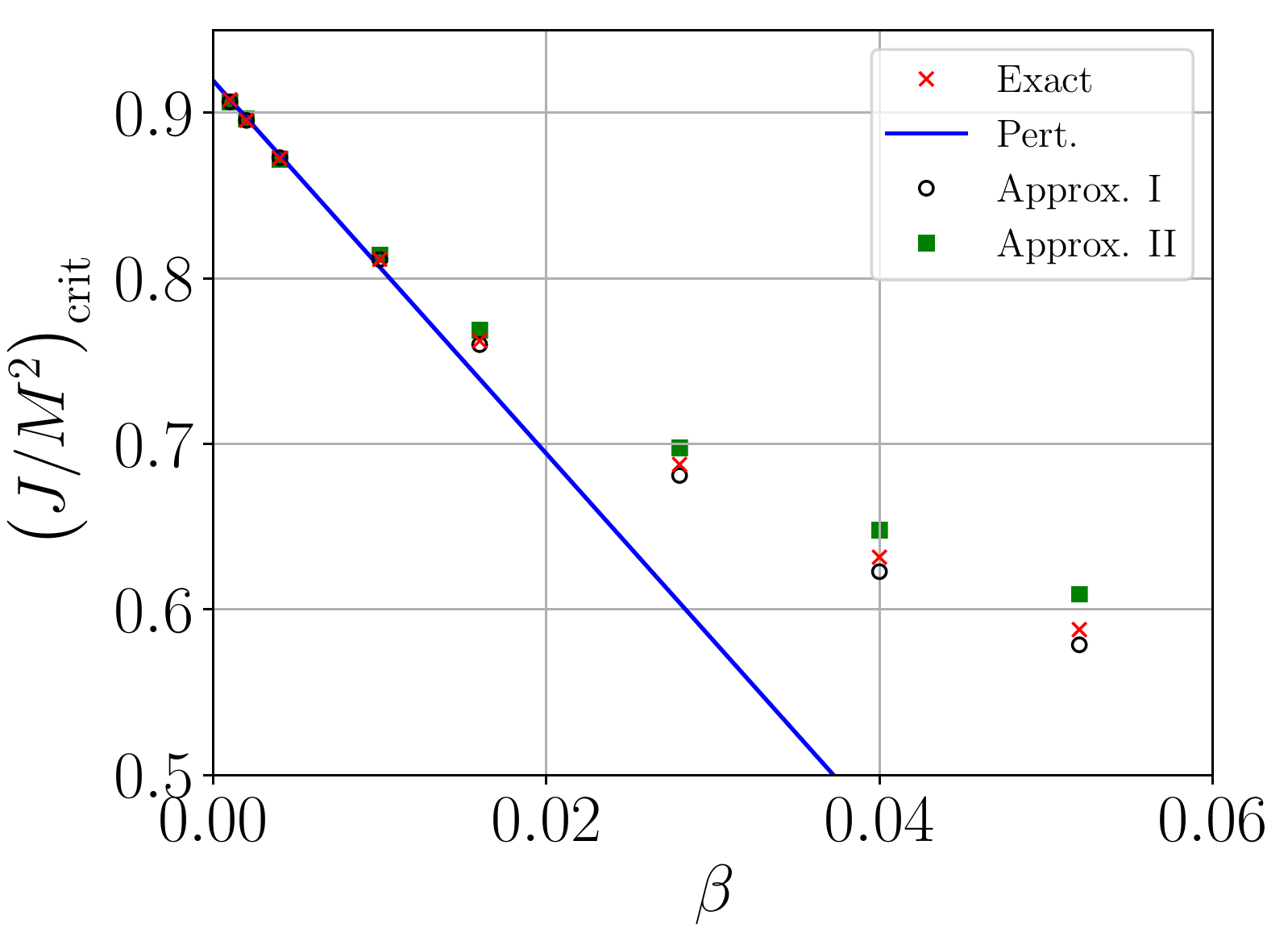}}}$

$\vcenter{\hbox{\includegraphics[width=0.47 \textwidth]{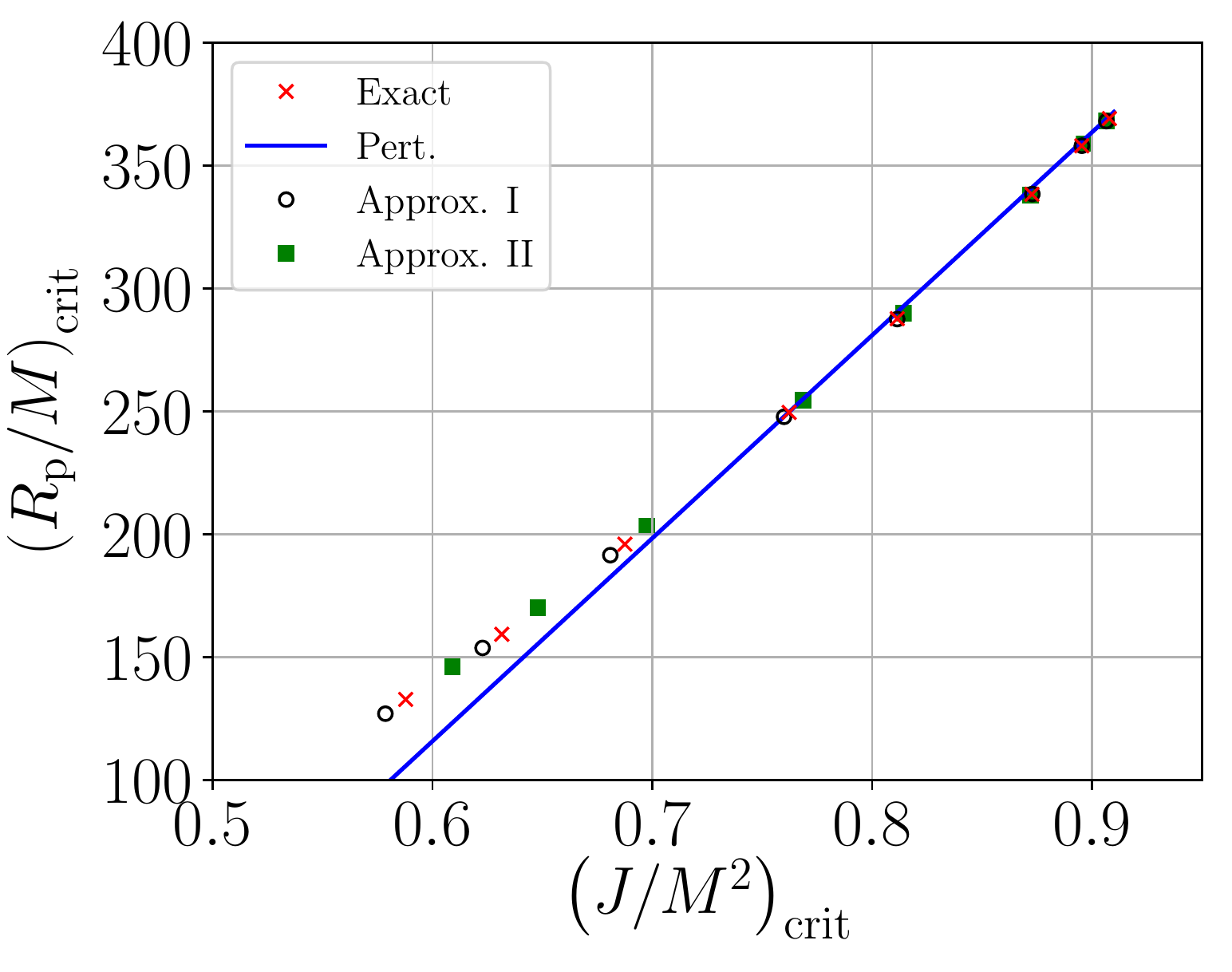}}}$
\end{center}
\caption{The dimensionless parameters $R_p/M$ and $J/M^2$ of the
  critical configuration, both as a function of $\beta$ (top left and
  top right panels) and plotted against each other (bottom panel).  As
  expected from Fig.~\ref{comparetofig3paperII}, our numerical results
  agree well with perturbative results for small values of $\beta$.
  For larger values of $\beta$, deviations between the exact treatment
  of the EOS and the two approximations increase as well, with
  Approximation I performing better than Approximation II.  When
  plotted against each other (bottom panel), values of $R_p/M$ and
  $J/M^2$ appear to lie on a single line, so that deviations in radius
  appear to be compensated for by deviations in the angular momentum.
  Note, however, that, according to different approaches, individual
  configurations on this line correspond to different values of
  $\beta$. }
\label{polarradandangmomfigbeta}
\end{figure*}

We summarize our results for critical configurations of maximally
rotating SMSs partially supported by gas pressure in
Figs.~\ref{polarradandangmomfigbeta} and \ref{polarradandangmomfig}.
In Fig.~\ref{polarradandangmomfigbeta} we show the dimensionless
parameters $R_p/M$ and $J/M^2$ as a function of $\beta$, as well as
plotted against each other, while in Fig.~\ref{polarradandangmomfig}
we show the parameters as a function of mass.  For the exact treatment
of the EOS we compute these physical masses by rescaling the
dimensionless masses $\bar M$ computed in the code according to $M =
K^{3/2} \bar M$, with $K$ given by (\ref{Kdef}).
\subsection{Approximation I}
\label{sec:rot:approachI}

\begin{figure*}
\begin{center}
\includegraphics[width=0.47 \textwidth]{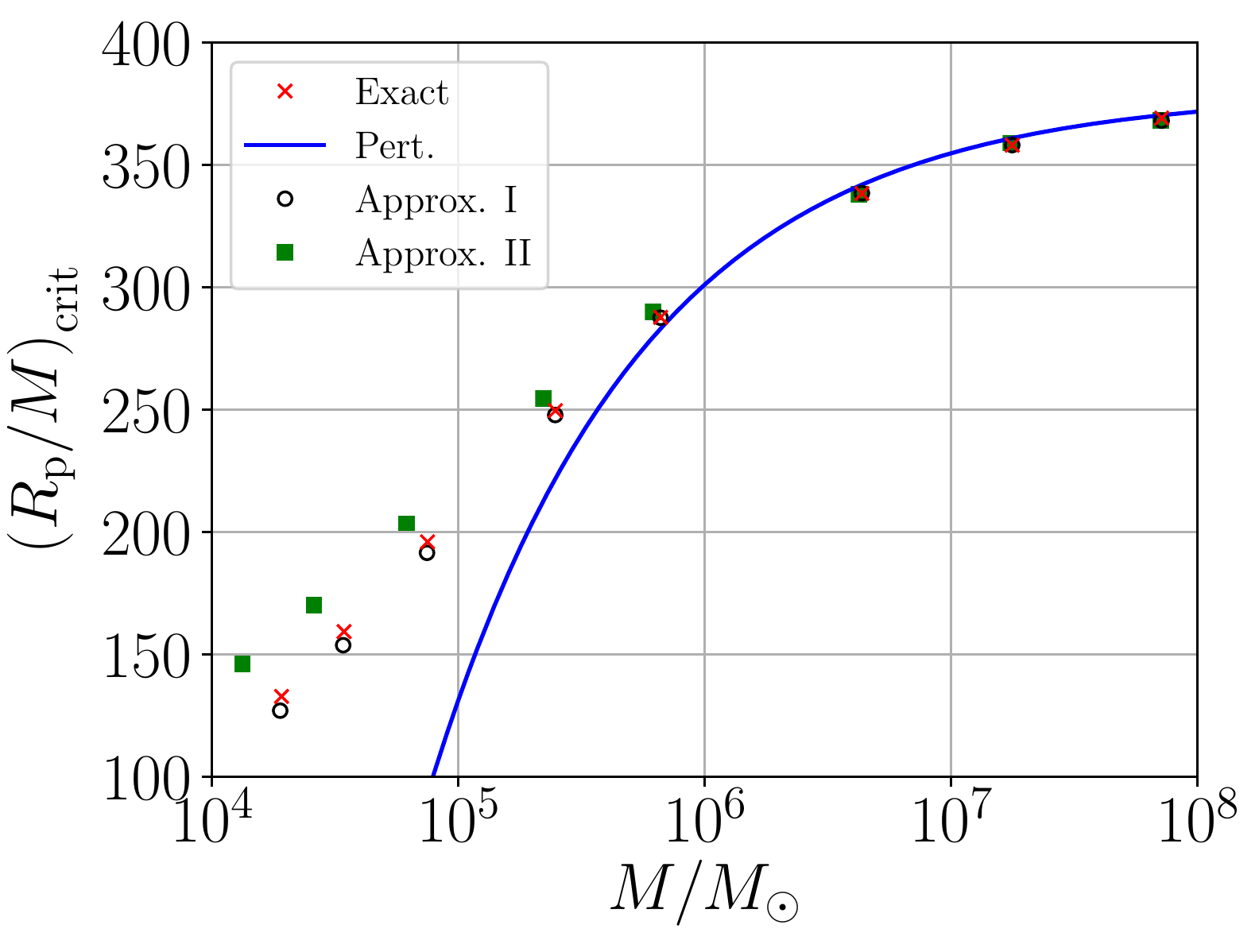}~~~
\includegraphics[width=0.47 \textwidth]{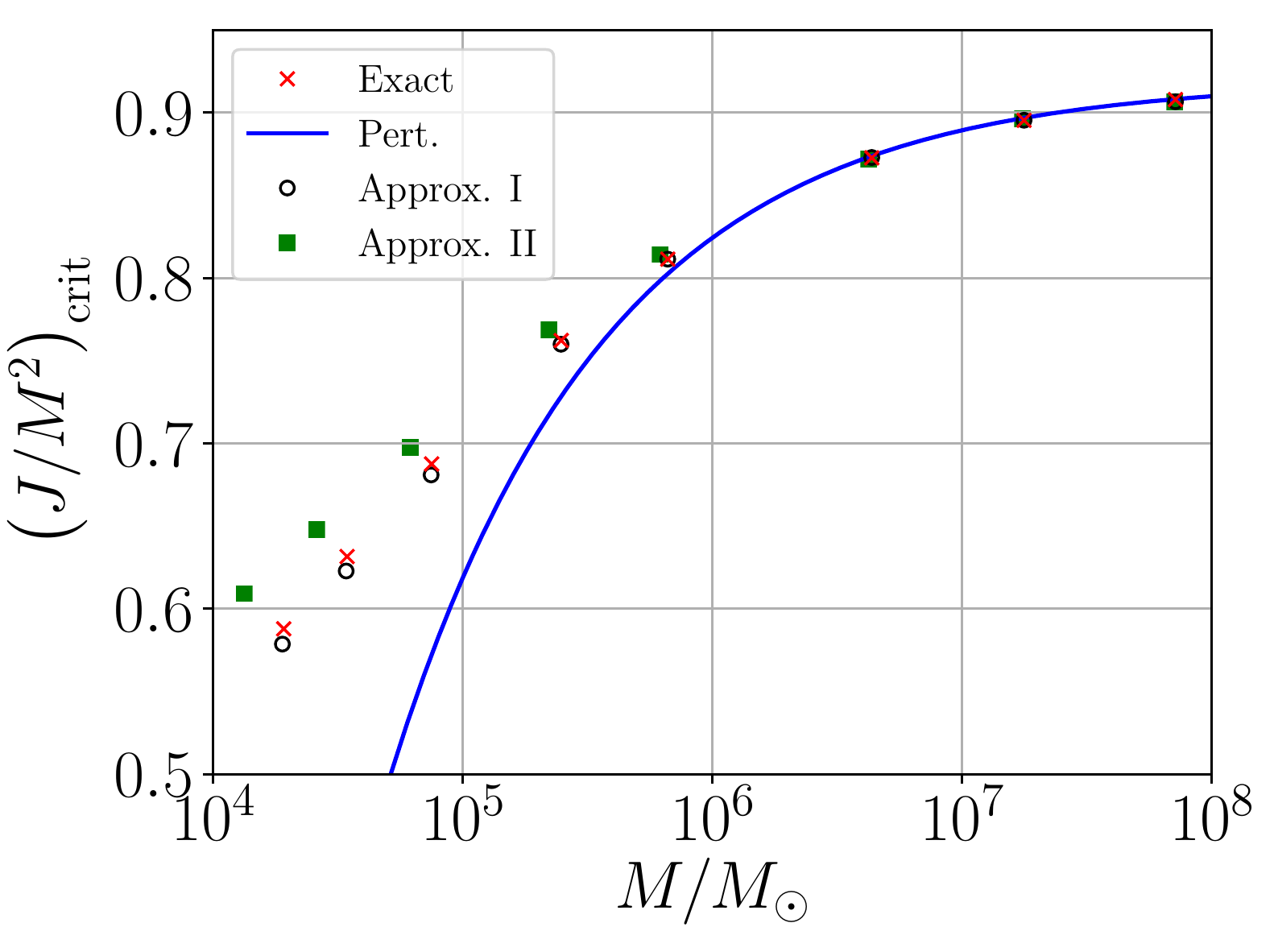}
\end{center}
\caption{Same as the top two panels in
  Fig.~\ref{polarradandangmomfigbeta}, expect plotted against the
  stellar mass $M$.  Compared to Fig.~\ref{polarradandangmomfigbeta},
  the differences between the numerical results appear to be slightly
  larger in this rendering, which is caused by differences in the
  rescaling of the mass (see text).}
\label{polarradandangmomfig}
\end{figure*}

For Approximation I we compute and analyze models of rotating SMS in
the same way as for the exact approach, except that we now run the RNS
code with EOS tables computed as discussed in Section
\ref{sec:eos:approachI}.  We show results from these calculations in
the middle column of Fig.~\ref{bigcomparisonfig}.  We again identify
the critical configurations for different values of $\beta$, compute
the corresponding changes from eqs.~(\ref{xrot}) ---(\ref{Mrot}), and
plot these changes in Fig.~\ref{comparetofig3paperII}.  We also graph
the parameters $R_p/M$ and $J/M^2$ in
Fig.~\ref{polarradandangmomfigbeta} and \ref{polarradandangmomfig},
where, for Approximation I, we have again computed the mass in
Fig.~\ref{polarradandangmomfig} from $M = K^{3/2} \bar M$, with $K$
given by (\ref{Kdef}).

In Paper II we adopted a perturbative approach within a simple energy
functional model to compute leading-order corrections to the critical
parameters.  Applying these methods for rotating SMSs and adopting
Approximation I, these changes are given by
\begin{equation}
\label{deltajrotI}
\delta_{j} = -\frac{1}{2}\delta{x} = -\frac{k_{1}}{8k_{3}}\frac{1}{\bar{M}_{0}^{2/3}\left(2j_{0}^{2} - j_{{\rm min}}^{2}\right)x_{0}}\beta
\end{equation}
(see (II.93)) and
\begin{equation}
\label{deltaMrotI}
\delta_{M}^{I} = \left(\frac{3}{4}\ln x_{0} + \frac{5}{4}\ln\beta + \frac{3}{2}C + \frac{9k_{5}}{4k_{3}}\frac{x_{0}}{2j_{0}^{2}-j_{{\rm min}}^{2}}\right)\beta,
\end{equation}
(see (II.96)), where $k_{3}=1.2041$ \citep{LaiRS93}, and
$k_{5}=0.331211$ (J.~C.~Lombardi Jr., 1997, priv.~comm.).  We also
find $j_{\rm min}=0.886$ for our $n=3$ polytrope simulations.  We use
these expressions to calculate the perturbative curves labeled
``Pert.~I" in Figure \ref{comparetofig3paperII}.  From $\delta_j$, we
can compute changes in the dimensionless ratios $R_p/M$ and $J/M^2$
from
\begin{equation}
\label{findRppert}
\left(\frac{R_{p}}{M}\right)_{\rm crit} = \left(\frac{R_{p}}{M}\right)_{\rm crit, 0}\left(1+2\delta_{j}\right)
\end{equation}
and
\begin{equation}
\label{findJpert}
\left(\frac{J}{M^{2}}\right)_{\rm crit} = \left(\frac{J}{M^{2}}\right)_{\rm crit, 0}\left(1+\delta_{j}\right)
\end{equation}
(see (II.87) and (II.88)).  These equations are plotted as the solid
lines in Figs.~\ref{polarradandangmomfigbeta} and
\ref{polarradandangmomfig}, using eq.~(\ref{deltajrotI}) and the
leading-order relationship between $\beta$ and $M$
\begin{equation}
\label{betaM}
\beta\approx 8.46\left(\frac{M}{M_{\sun}}\right)^{-1/2}
\end{equation} 
(see, e.g., (II.40)).

\subsection{Approximation II}
\label{sec:rot:approachII}

Finally we repeat the same exercise with EOS tables computed from
Approximation II, as discussed in Section \ref{sec:eos:approachII}.
We show numerical results in the right column of
Fig.~\ref{bigcomparisonfig}.  As before, critical configurations are
marked by red dots.  We identify the physical parameters for these
critical configurations, compute changes from eqs.~(\ref{xrot})
-- (\ref{Mrot}), and plot these changes in
Fig.~\ref{comparetofig3paperII}.  As before, we also graph the polar
radius and the angular momentum in
Figs.~\ref{polarradandangmomfigbeta} and \ref{polarradandangmomfig}.
In the latter, we now compute the mass from $M = K_{II}^{n_1/2} \bar
M$, with $K_{II} \approx (1 + \beta) K$ and $n_1$ given by
Eq.~(\ref{n1App2}) (see Section \ref{sec:eos:approachII}).

Adopting Approximation II in the perturbative treatment of the energy
functional approach leads to the same $\delta_j$ as Approximation I,
given by (\ref{deltajrotI}), while $\delta_M$ is now given by
\begin{equation}
\label{deltaMrotII}
\delta_{M}^{II} = \left(\frac{3}{4}-\frac{1}{2}\ln\bar{M}_{0}+\frac{3}{4}\ln x_{0}-\frac{3}{4}\frac{j_{0}^{2}}{2j_{0}^{2}-j_{{\rm min}}^{2}}\right)\beta.
\end{equation}
(see (II.105)).  We use these expressions to calculate the perturbative curves labeled ``Pert.~II" in Fig.~\ref{comparetofig3paperII}.   Since expressions for $R_p/M$ and $J/M^2$ are the same in Approximation I and II, both are represented by the same perturbative line in Figs.~\ref{polarradandangmomfigbeta} and \ref{polarradandangmomfig}.

\subsection{Comparison}
\label{sec:rot:comparison}

Figs.~\ref{comparetofig3paperII}, \ref{polarradandangmomfigbeta}, and
\ref{polarradandangmomfig} show that, for small $\beta$,
corresponding to large masses, all approaches, including the
perturbative treatment, lead to similar predictions for dimensionless
quantities, including the dimensionless parameters $R_p/M$ and $J/M^2$
characterizing the critical configuration.  In particular, our
numerical results confirm our perturbative finding of Paper II that,
even for masses as large as $M \simeq 10^6 M_{\odot}$, gas pressure
has a significant effect on these parameters.  The reason for this
behavior is the fact that, to leading order, corrections to the
parameters scale with $M^{-1/2}$, and therefore decrease only slowly
as the mass increases (see Eqs.~(II.142) and (II.142)).  For moderate
values of $\beta$, or stellar masses $\lesssim 10^6 M_{\odot}$, the 
analytic perturbative treatment
starts to deviate from the numerical results, while both approximations 
implemented numerically continue to agree
with each other up to larger values of $\beta$, and smaller masses.
Ultimately, Approximation II in particular shows increasing deviations
from the exact treatment as well.

As we had noted in Paper II, Approximation II results in predictions
for changes in the mass that differ from Approximation I and the exact
treatment even at leading order, both in the numerical and
perturbative treatments (see the right panel in
Fig.~\ref{comparetofig3paperII}).  We believe that at least some of
these deviations may be related to the different scaling used for the
different approaches: for the exact treatment and Approximation I, we
rescale dimensional quantities with $K^{3/2}$ (see Eq.~(\ref{Kdef})),
while for Approximation II we rescale with $K_{II}^{n_1/2}$.  Since
$K_{II}$ is only approximately constant, this approximation may well
be responsible for deviations that we find in dimensional quantities.

In the left two panels of Fig.~\ref{polarradandangmomfigbeta} we show
the dimensionless parameters $R_p/M$ and $J/M^2$ as a function of
$\beta$.  As before, we find that, as the gas pressure becomes more
important, Approximation II leads to larger deviations from the exact
treatment than Approximation I.  In the right panel of
Fig.~\ref{polarradandangmomfigbeta} we plot $R_p/M$ versus $J/M^2$.
Remarkably, all three approaches lead to results that appear to follow
a single line, even though, according to the different treatments of
the gas pressure, individual configurations on this line would be
identified with different values of $\beta$.  Finally we graph $R_p/M$
and $J/M^2$ against stellar masses $M$ in
Fig.~\ref{polarradandangmomfig}.  The deviations between the exact
treatment of the EOS and Approximation II now appear slightly larger
than in the left two panels of \ref{polarradandangmomfig}, which we
attribute to the scaling of the mass, as discussed above.

%
\section{Summary and Discussion}
\label{sec:sum}
%


Recent observations of increasingly young quasars 
have heightened interest in the direct-collapse
scenario for the formation of SMBHs, in which a SMS becomes unstable
and collapses gravitationally.  
A number of groups have studied possible avenues for the formation of SMSs
\citep[see, e.g.,][see also the discussion in Section \ref{sec:intro}]{SchPFGL13,HosYIOY13,SakHYY15,UmeHOY16,WooHWHK17,HaeWKHW18a,HaeWKHW18b,WisRONDX19}. 
In this paper we continue the study of
an idealized version of a direct-collapse scenario, involving uniformly 
rotating SMSs evolving along the mass-shedding limit until they reach a critical
configuration marking the onset of radial instability (see Papers I and II).  Identifying this critical configuration
is important since it determines the dynamics of the subsequent collapse to a SMBH, including the 
accompanying gravitational wave signal and the properties of the remnant.  In fact, many fully
relativistic simulations of this collapse have adopted models of the critical configuration
as initial data \citep[see, e.g.][]{ShaS02,LiuSS07,MonJM12,ShiSUU16,UchSYSU17,SunRS19}.
In this paper we study the effects of gas pressure on the critical configuration.  While we believe
that our findings are interesting in their own right, we also hope that they will help improve future
dynamical simulations of the collapse of SMSs to SMBHs.

In Paper I we
found that the critical configuration is characterized by unique
values of $R_p/M$ and $J/M^2$ as long as the star is dominated by
radiation pressure.  In Paper II we computed leading-order corrections
to these values when some of the assumptions of Paper I were relaxed;
in particular we considered two different approximations to estimate
the effects of gas pressure.  Approximation I is based on a formal
expansion, while Approximation II accounts for the effects of gas
pressure by simply adjusting the polytropic index in a polytropic EOS.
The latter is therefore simple to implement and has been used quite
commonly.  Somewhat surprisingly, we found that some predictions
stemming from these two approximations differed.

In this paper we apply the turning-point criterion to study 
more systematically the effects of gas pressure
on the critical configuration of maximally rotating SMSs, and
determine the critical configuration and its parameters for a large
range of stellar masses.  We also evaluate differences stemming from
different treatments of the gas pressure.  To do so, we expand on our
treatment in Paper II in two ways.  Instead of employing a
perturbative analysis within a simple analytic energy functional model, we now
compute fully relativistic numerical models of rotating SMSs.  We also
include a fully self-consistent, exact treatment of the EOS, in
addition to the two approximations discussed above, so that we can
calibrate the two approximations in the context of this idealized
direct-collapse scenario.

As expected, all methods agree well for large masses, $M \gtrsim 10^6
M_{\odot}$, corresponding to large entropies, and hence to small
$\beta$ and small effects of the gas pressure.  In particular, our
numerical results confirm the perturbative results of Paper II that,
even for these large masses, the effects of gas pressure are
important.  Not surprisingly, the perturbative treatment starts to
deviate from the exact results first as $\beta$ increases and the mass
decreases.  Below $M \simeq 10^5 M_{\odot}$, both approximations
lead to increasing deviations from the exact treatment of gas pressure, 
but Approximation I
remains much closer to the exact results than Approximation II.

\section*{Acknowledgments}

This work was supported in part by NSF grant PHYS-1707526 to Bowdoin
College, NSF grant 
PHY-1662211 and NASA grant 80NSSC17K0070 to the
University of Illinois at Urbana-Champaign, as well as through
sabbatical support from the Simons Foundation (Grant No.~561147 to
TWB).

\bibliographystyle{mnras}

\label{lastpage}
\end{document}